\documentclass[extra,9pt]{gji}  
\onecolumn
\usepackage[font={small}]{caption}   
\usepackage[fleqn]{amsmath}
\usepackage{setspace,url}
\usepackage{mathrsfs}  
\usepackage{times}
\usepackage{graphicx}
\usepackage{listings}
\usepackage{enumerate}
\usepackage{lineno}
\usepackage{mathtools}
\newcommand{\model}{0.3\columnwidth}
\newcommand{\traces}{0.75\columnwidth}
\newcommand{\kernels}{0.35\columnwidth}
\newcommand{\arrays}{0.374\columnwidth}
\newcommand{\arrayss}{0.3\columnwidth}
\renewcommand{\models}{0.575\columnwidth}

\usepackage{xspace}
\newcommand{\cT}{\mathcal{T}}
\newcommand{\sij}{_{ij}}
\newcommand{\sji}{_{ji}}
\newcommand{\sjj}{_{jj}}

\newcommand{\syn}{^{\mathrm{syn}}}
\newcommand{\obs}{^{\mathrm{obs}}}
\newcommand{\dtss}{\Delta t\sij\syn}
\newcommand{\dtso}{\Delta t\sij\obs}
\newcommand{\dtssA}{\Delta t_{12}\syn}
\newcommand{\dtsoB}{\Delta t_{12}\obs}
\newcommand{\ddtN}{\frac{\dd t\sij}{N\sij}}
\newcommand{\DlnA}{\Delta\!\ln\!{A}}
\newcommand{\dDlnA}{\delta\DlnA}
\newcommand{\DDlnA}{\Delta\DlnA}
\newcommand{\fracd}[2]{\frac{\displaystyle{#1}}{\displaystyle{#2}}}
\newcommand{\ofo}{(\omega)}
\newcommand{\oft}{(t)}
\newcommand{\oftt}{(T-t)}
\newcommand{\ofs}{}
\newcommand{\hspso}{\hspace{0.1em}}
\newcommand{\dt}{\hspace{0.1em}\mathrm{d}t}
\newcommand{\dtt}{\mathrm{d}t}

\newcommand{\dx}{\hspace{0.1em}\mathrm{d}^3\bx}
\newcommand{\intot}{\int_0^T}
\newcommand{\pl}{\partial}
\newcommand{\PP}{_\Phi}
\newcommand{\A}{_A}
\newcommand{\CC}{_{\tiny \mathrm{cc}}}
\newcommand{\DD}{^{\tiny \mathrm{dd}}}
\newcommand{\dd}{\Delta\Delta} 
\newcommand{\also}{\qquad\mbox{and}\qquad}
\newcommand{\where}{\quad\mbox{where}\quad}
\newcommand{\with}{\quad\mbox{with}\quad}
\newcommand{\Tfs}{T\sij\syn}
\newcommand{\cTfs}{\cT\sij\syn}
\newcommand{\half}{\frac{1}{2}}
\newcommand{\hsps}{\hspace{0.75em}}

\newcommand{\edc}{\end{document}}

\newcommand{\bx}{\mathbf{x}}

\newcommand{\var}{\mbox{var}}
\renewcommand\Re{\operatorname{Re}}
\renewcommand\Im{\operatorname{Im}}
\newcommand{\atan}{\mbox{atan}}

\newcommand{\sumh}{\sum_{i}\sum_{j>i}}
\newcommand{\sumk}{\sum_k^K}
\newcommand{\suml}{\sum_l^K}
\newcommand{\sumo}{\sum_\omega}
\DeclareMathOperator*{\argmin}{\arg\!\min}
\DeclareMathOperator*{\argmax}{\arg\!\max}

\begin{document}
\title[Double-difference adjoint seismic tomography]{Double-difference
  adjoint seismic tomography}
\renewcommand{\thefootnote}{\fnsymbol{footnote}} \author[Y.O.~Yuan, F.J.~Simons
  and J.~Tromp]{Yanhua O.~Yuan, Frederik J.~Simons, and Jeroen
  Tromp\\ Department of Geosciences, Princeton University, Princeton,
  NJ 08544, USA\\E-mail: yanhuay@princeton.edu} \maketitle
\pubyear{2016}

\begin{summary}
  We introduce a `double-difference' method for the inversion for
  seismic wavespeed structure based on adjoint tomography. Differences
  between seismic observations and model predictions at individual
  stations may arise from factors other than structural heterogeneity,
  such as errors in the assumed source-time function, inaccurate
  timings, and systematic uncertainties. To alleviate the
  corresponding nonuniqueness in the inverse problem, we construct
  differential measurements between stations, thereby reducing the
  influence of the source signature and systematic errors. We minimize
  the discrepancy between observations and simulations in terms of the
  differential measurements made on station pairs. We show how to
  implement the double-difference concept in adjoint tomography, both
  theoretically and in practice. We compare the sensitivities of
  absolute and differential measurements. The former provide absolute
  information on structure along the ray paths between stations and
  sources, whereas the latter explain relative (and thus
  higher-resolution) structural variations in areas close to the
  stations. Whereas in conventional tomography a measurement made on a
  single earthquake-station pair provides very limited structural
  information, in double-difference tomography one earthquake can
  actually resolve significant details of the structure. The
  double-difference methodology can be incorporated into the usual
  adjoint tomography workflow by simply pairing up all conventional
  measurements; the computational cost of the necessary adjoint
  simulations is largely unaffected. Rather than adding to the
  computational burden, the inversion of double-difference
  measurements merely modifies the construction of the adjoint sources
  for data assimilation.\\[1em]
\end{summary}

\section{\label{introduction}I~N~T~R~O~D~U~C~T~I~O~N}

The quality of tomographic inversions for seismic wavespeed structure
is affected by uncertainties in our knowledge of the source-time
function (the `source wavelet' in exploration seismology), the
(earthquake) source mechanism, location and origin time, inaccuracies
in record timekeeping, network geometry, and systematic uncertainties
\cite[]{Nolet2008}.  All of these factors introduce intrinsic errors
into synthetic modeling, even when accurate information on structure
is available. Seismic tomography --- in seeking to maximize the
agreement between simulations and observations from all sources at
all recording stations --- is at risk of mapping errors of this kind
into estimates of wavespeed variations
\cite[]{Schaeffer+2013}. Jointly updating source terms and structural
model parameters \cite[]{Pavlis+80,Spencer+80,Abers+91,Thurber92},
sequentially or iteratively, is a workable albeit expensive solution
\cite[]{Widiyantoro+2000,Panning+2006b,Tian+2011,Simmons+2012}.

The fundamental ideas of making differential measurements were
developed early on \cite[e.g.][]{Brune+63,Passier+95b}.  The
earthquake location community pioneered the concept of
`double-difference' inversions \cite[]{Poupinet+84,Got+94}, and a
popular code, \texttt{hypoDD}, was introduced by
\cite{Waldhauser+2000} to improve the accuracy of source locations.
The double-difference tomography code \texttt{tomoDD}
\cite[]{Zhang+2003b} produces high-resolution wavespeed models
simultaneously with accurate event locations. Rather than considering
pairs of earthquakes recorded at each station to reduce the effects of
structural uncertainty on the source locations, differencing
differential measurements between pairs of stations recording the same
earthquake lessens the effect of uncertainties in the source terms on
the determination of Earth structure
\cite[][]{Monteiller+2005,Fang+2014}. Much as the double-difference
technique has revolutionized earthquake (and other source) studies
\cite[e.g.][]{Rubin+99,Rietbrock+2004,Schaff+2004b}, remaining the
method of choice for high-resolution hypocenter determination, seismic
tomography gains from fully embracing those concepts for structural
inversions.

Until now, most of the double-difference approaches have operated
within the confines of ray theory \cite[but
  see][]{deVos+2013,Maupin+2015}. In this paper, we bring
double-difference inversions to full-waveform adjoint seismic
tomography. Finite-frequency traveltime measurements drive a nonlinear
inversion strategy that performs misfit gradient calculations via an
elastic adjoint-state method. \cite{Dahlen+2000} and \cite{Hung+2000}
approximated the Fr\'{e}chet kernel for differential traveltimes,
measured by cross-correlation of phases (e.g., \textit{P}, \textit{S},
\textit{pP}, \textit{SS}, ...)  with `identical' pulse shapes, by the
difference of the individual single-phase kernels calculated
asymptotically. \cite{Hung+2004} used the same approximation to
measure the sensitivities of relative time delays between two nearby
stations, which showed improved resolution of the upper-mantle
structure beneath a regional array. \cite{Tromp+2005} brought the
elastic adjoint method of \cite{Tarantola87b,Tarantola88} into global
seismology, calculating sensitivity kernels accurately and efficiently
using a spectral-element method
\cite[]{Komatitsch+2002a,Komatitsch+2002b}. With these techniques it
is now possible to incorporate essentially any type of seismic
`measurement' (absolute, relative, or differential) into global
inversions for seismic wavespeed, boundary perturbations, and source
parameters. In this paper, we derive a constructive theory for the
incorporation of generic `double-difference' measurements into adjoint
seismic tomography.

The minimization of the difference of the difference between
observations at distinct pairs of stations, and the difference between
synthetics at the same pairs, over all station pairs, requires
explicit mathematical expressions for the adjoint sources that are
necessary for the numerical computation of the corresponding misfit
gradient functionals. We derive those and show in numerical
experiments that the gradients of the new misfit functional with
respect to structural perturbations are relatively insensitive to an
incorrect source signature and timing errors. These results stand in
contrast to conventional tomography, which aims to minimize the
difference between predictions and observations obtained from
measurements made at individual stations.

A seismic phase observed at two `nearby' (relative to the average
source-receiver distance) stations will have `similar' (using robust
metrics) waveforms and `sensitivities' (misfit gradients) to Earth
structure. Station-relative differential measurements will reflect
smaller-scale structural variations near to and in-between the station
pairs. Structural inversions under the conventional formalism, unless
actively `re'- or `pre'-conditioned to avoid doing so
\citep[]{Curtis+97,Spakman+2001,Fichtner+2011b,Luo+2015}, tend to
over-emphasize areas with high-density ray-path coverage relative to
poorly sampled parts of the model.  By the partial cancellation of
common sensitivities, double-difference tomography, on the contrary,
illuminates areas of the model domain where ray paths are not densely
overlapping. Hence, using only one event (e.g., a teleseismic
earthquake) recorded at a cluster of stations, conventional tomography
can derive only very limited structural information, while
double-difference tomography will resolve the structure in the
instrumented area in greater detail, a benefit that accrues with the
density of seismometer arrays \cite[]{Rost+2002,Burdick+2014}.

The computational cost for tomographic inversion is essentially
unaffected by incorporating the double-difference concept into the
conventional adjoint tomography workflow. All existing individual
measurements are simply paired up, and only the construction of the
adjoint sources requires modification. Data assimilation happens by
back-propagating all the resulting adjoint sources simultaneously to
compute the gradient kernels, exactly as for conventional adjoint
tomography, without additional computational burden. Furthermore, the
differential measurements themselves can be composed from any of the
types used in conventional tomography. For example, they can be
relative times calculated from a catalog of absolute arrival times
\cite[e.g.,][]{VandeCar+90}; they may relative traveltime delays
obtained from waveform cross-correlation analysis
\cite[e.g.,][]{Luo+91}, or from the cross-correlation of their
envelopes \cite[e.g.,][]{Yuan+2015}, and so on. The comparison need
not be between the same phase observed at two distinct stations; it
can involve different phases recorded on the same trace, or signals
recorded at the same station but originating from two different
events. In this paper, we use common waveform cross-correlation
traveltimes as examples of incorporating differential measurements
into adjoint-based tomography. For dispersive waves such as surface
waves, differential frequency-dependent phase and amplitude
measurements can be made using multitaper cross-spectral analysis. The
details of that procedure are relegated to the Appendix.

Double-difference adjoint tomography relies on the assimilation of
measurements, which we discuss how to group (e.g., via cluster
analysis), and how to weight and stabilize (e.g., by regularization).
We propose two approaches to cluster analysis and regularization. The
first is based on source-station geometry, in which only station pairs
whose spacing is comparable to the scale of the resolvable wavelengths
are included in the double-difference data ensemble. The second is
based on waveform similarity, as evaluated by the maximum normalized
cross-correlation between two waveforms. When the waveform similarity
exceeds a certain threshold, the pair of stations is integrated in the
double-difference group, and the relative contributions to the misfit
function from all qualifying station pairs are weighted by their
similarity.

We demonstrate how to make double-difference measurements, and how to
use them for adjoint tomography. Numerical experiments show the
sensitivities of the double-difference data compared with conventional
absolute measurements. We conduct tests with realistic network
configurations, both on a global scale and at the scale of the North
American continent, and for wavespeed structures that are either
checkerboard synthetics or plausible Rayleigh-wave phase speed maps
inspired by prior studies.

\section{A~D~J~O~I~N~T{\hsps}T~O~M~O~G~R~A~P~H~Y~:{\hsps}T~H~E{\hsps}C~L~A~S~S~I~C~A~L{\hsps}A~P~P~R~O~A~C~H} 

We briefly review the principles of `conventional' adjoint tomography
using `absolute' cross-correlation traveltimes.  The material in this
section is later used for comparison with the `double-difference'
adjoint tomography using `differential' cross-correlation traveltimes.

The cross-correlation traveltime difference between a synthetic signal
$s_i(t)$ and an observation $d_i(t)$ over a window of length~$T$ is
defined as
\begin{align}\label{eq:cc_sd}
  \Delta t_i&=\argmax_\tau \intot s_i(t+\tau) d_i(t) \dt
  ,
  \qquad
  \mbox{and we define}
  \qquad
  \gamma_{i}=\intot s_i(t+\tau) d_i(t) \dt
  .
\end{align}
A positive $\Delta t_i$ indicates that the data waveform~$d_i$ is
advanced relative to the synthetic~$s_i$, meaning that the wavespeed
model that generated the synthetics is slower than the true model. The
objective function to be minimized by `least-squares' is the sum over
all measurements of the squared traveltime shifts in
eq.~(\ref{eq:cc_sd}):
\begin{align}\label{eq:misfit_CC}
  \chi\CC&=\half\sum_i  [\Delta t_i]^2
  .
\end{align}
Explicit and complete derivations for the Fr\'echet derivatives of the
terms in the cross-correlation traveltime misfit function in
eq.~(\ref{eq:misfit_CC}) were presented by various authors
\cite[e.g.,][]{Luo+91,Marquering+99,Dahlen+2000}. Most central to the
development,
\begin{equation}\label{dti}
  \delta\Delta t_i=
  \frac{\intot
    \pl_t s_i(t)\,
    \delta s_i(t)\dt
  }
       {\intot \pl^2_t s_i(t)  s_i(t) \dt}
       .
\end{equation}
To turn the expression for the traveltime perturbation,
eq.~(\ref{dti}), into a useful expression for the perturbation of the
misfit function in eq.~(\ref{eq:misfit_CC}) requires a mechanism to
relate a change in Earth properties (density and elastic constants or
wavespeeds) to a change in the seismogram, $\delta s_i(t)$. Over the
years, various formalisms were developed (acoustic, ray-based,
modes-based, and numerical approaches), and they were amply discussed
in the literature. In addition to the works cited above, we must still
mention \cite{Zhao+2000}, \cite{Chen+2007b} and
\cite{Nissen-Meyer+2007a}. Here we follow \cite{Tromp+2005} in taking
the numerical approach, which involves the action of an `adjoint
source'. For an individual measurement of cross-correlation traveltime
made at~$\bx_i$, the adjoint source
\begin{align}\label{eq:adj_CC}
  f_i^{\dagger}(\bx,t)&= 
  \Delta t_i
  \frac{\pl_t s_i(T-t)}
       {\intot \pl^2_t s_i(t) \, s_i(t) \dt} \, \delta (\bx-\bx_i)
       .
\end{align}
Note the weighting by the traveltime anomalies~$\Delta t_i$.  The
reverse-time synthetics generated at each of the stations from the
corresponding adjoint sources in eq.~({\ref{eq:adj_CC}}) are summed
and simultaneously back-propagated. The interaction of the adjoint
with the forward-propagating wavefield then produces the gradients of
the objective function in eq.~(\ref{eq:misfit_CC}) with respect to the
parameterized model perturbations~$m$, leading to expressions that
embody the essence of the inverse problem in seismic tomography
\cite[]{Nolet96}, namely
\begin{equation}\label{kp}
  \delta\chi\CC=\int_\oplus K_m(\bx) \,m(\bx)\dx
  ,
  \where
  K_m(\bx)~\mbox{is the cross-correlation traveltime misfit sensitivity kernel for parameter}~m
  .
\end{equation}
Positive cross-correlation traveltime differences $\Delta t_i$ from
eq.~(\ref{eq:cc_sd}) result in negative kernel values~$K_m$ in
eq.~(\ref{kp}). The model update required is in the direction
opposite to~$K_m$, which, for positive~$\Delta t_i$ and
negative~$K_m$, implies that the model wavespeeds need to be sped up
to reduce the misfit in eq.~(\ref{eq:misfit_CC}). The integrations
are carried out over the entire volume of the Earth, and the
summation over the set of parameters~$m$ is implied.  The typical
isotropic situation would be for eq.~(\ref{kp}) to involve the
density and elastic moduli $\rho(\bx)$, $\kappa(\bx)$ and $\mu(\bx)$,
or the density and the elastic wavespeeds $\rho(\bx)$, $\alpha(\bx)$
and $\beta(\bx)$, and so on. \cite{Tape+2007}, \cite{Fichtner+2008},
\cite{Zhu+2009}, and others, describe in detail how the forward and
adjoint wavefields interact to produce the various types of `misfit
sensitivity kernels' under different model parameterizations, and for
the specific case considered in this section. An excellent recent
overview is by \cite{Luo+2015}.

\section{A~D~J~O~I~N~T{\hsps}T~O~M~O~G~R~A~P~H~Y~:{\hsps}T~H~E{\hsps}D~O~U~B~L~E~-~D~I~F~F~E~R~E~N~C~E{\hsps}W~A~Y}

We use the term `double-difference' measurement for the difference
(between synthetics and observations) of differential measurements
(either between pairs of stations or between pairs of
events). From the new measurement we construct a new misfit
function and derive its Fr\'echet derivatives and adjoint sources. In
this section we focus on `measurements' made by cross-correlation, and
take `differential' to mean `between pairs of stations, from a common
source'. Our definitions can be relaxed later to apply to other types
of measurements (e.g., waveform differences, envelope differences), and
to refer more broadly to differential measurements, e.g., between
different seismic phases or between different seismic sources observed
at the same station.

\subsection{Measurement}

We consider the case of differential traveltimes calculated by
cross-correlation of waveforms from a common source recorded at a pair
of stations indexed $i$ and $j$. Let $s_i(t)$ and $s_j(t)$ denote a
pair of synthetic waveforms, and let $d_i(t)$ and $d_j(t)$ be the
corresponding pair of observations. The differential cross-correlation
traveltimes, between stations, for the synthetic and the observation
pairs are, respectively,
\begin{subequations}
  \label{eq:cc}
  \begin{align}
    \label{eq:cc_ss}
    \dtss&=\argmax_\tau  \intot s_i(t+\tau)s_j(t) \dt
    ,
    \qquad
    \mbox{and we define}
    \qquad
    \Gamma\sij(\tau)=  \intot s_i(t+\tau)s_j(t) \dt,
    \\
    \label{eq:cc_dd}
    \dtso&=\argmax_\tau  \intot d_i(t+\tau)d_j(t) \dt
    ,
    \qquad
    \mbox{and we define}
    \qquad
    \Lambda\sij(\tau)= \intot d_i(t+\tau)d_j(t) \dt.
  \end{align}
\end{subequations}
A positive $\Delta t_{ij}$ indicates that the waveform recorded at
station~$j$ (i.e., $s_j$ or $d_j$) is advanced relative to the
waveform recorded at station~$i$ (i.e., $s_i$ or $d_i$). The difference of
these differential traveltimes, between synthetics and observations,
is the `double-difference' traveltime measurement:
\begin{align}\label{eq:cc_DD}
  \dd t\sij &= \dtss -\dtso.
\end{align} 
A positive $\dd t\sij$ indicates that the advancement of $s_j$ over $s_i$
is larger than that of $d_j$ over $d_i$.

\subsection{Misfit function}

The misfit that we minimize is the sum of squares of the
double-difference traveltime measurements, from eq.~(\ref{eq:cc_DD}),
between all station pairs:
\begin{align}\label{eq:misfit_CC_DD}
  \chi\CC\DD&=\half\sumh [\dd t\sij]^2
  .
\end{align}
The summation avoids double counting from the symmetry of the
traveltime measurements, since $\Delta t\sji\syn = -\dtss$
and $\Delta t\sji\obs = -\dtso$.

\subsection{Misfit gradient} 

The derivative of the differential objective function in
eq.~(\ref{eq:misfit_CC_DD}) is
\begin{align}\label{eq:derivative_misfit_CC_DD}
  \delta \chi\CC\DD&= \sumh[\dd t\sij]\,\delta\dtss,
\end{align}
where $\delta\dtss$ is the perturbation to the differential
traveltime synthetic~$\dtss$ caused by a model
perturbation. As is customary we approximate the synthetic wavefield in the
perturbed model to first order as the sum of the unperturbed
wavefield~$s(t)$ and a perturbed wavefield~$\delta s(t)$,
\begin{equation}\label{eq:wavefield}
  \tilde{s}_i(t)=s_i(t)+\delta s_i(t) \also
  \tilde{s}_j(t)=s_j(t)+\delta s_j(t).
\end{equation}
Hence, from eqs~(\ref{eq:cc_ss}) and~(\ref{eq:wavefield}), the
cross-correlogram of the new seismograms $\tilde{s}_i(t)$ and
$\tilde{s}_j(t)$ is, to first order in the perturbation,
\begin{align}\label{eq:Gamma}
  \tilde{\Gamma}\sij(\tau)
  &= \intot\tilde{s}_i(t+\tau)\tilde{s}_j(t)\dt
  \approx \Gamma\sij(\tau) + \intot \delta s_i(t+\tau) s_j(t)\dt +
  \intot s_i(t+\tau) \delta s_j(t)\dt
  = \Gamma\sij(\tau) + \delta \Gamma_i(\tau) + \delta\Gamma_j(\tau)
  .
\end{align}
We introduced the notation $\delta \Gamma_i(\tau)$ for the
cross-correlogram of $\delta s_i(t)$ an $s_j(t)$, and $\delta
\Gamma_j(\tau)$ for that between $s_i(t)$ and $\delta s_j(t)$, noting that
\begin{align}
  \delta \Gamma_i(\tau) 
  &=\intot \delta s_i(t+\tau)s_j(t)\dt= \intot s_j(t-\tau) \delta s_i(t)\dt
  \also
  \delta  \Gamma_j(\tau) = \intot s_i(t+\tau) \delta s_j(t)\dt. 
  \label{eq:Gamma_part3}
\end{align}
We recall from the definition of the differential cross-correlation
traveltime in eq.~(\ref{eq:cc_ss}) that the unperturbed
cross-correlogram $\Gamma\sij(\tau)$ achieves its maximum at
$\tau=\dtss$, in other words, $\pl_\tau \Gamma\sij(\dtss) = 0 $.  If
we expand the perturbed cross-correlogram $\tilde{\Gamma}\sij(\tau)$
in the vicinity of the unperturbed cross-correlation maximum
differential time $\dtss$, keeping terms up to second order, we obtain
\begin{align}\label{eq:Gamma_expand}
  \tilde{\Gamma}\sij (\dtss+\delta\tau)
  &= \Gamma\sij(\dtss)
  +\delta \Gamma_i(\dtss) 
  +\delta \Gamma_j(\dtss) 
  +\delta\tau \, \pl_\tau \delta \Gamma_i(\dtss) 
  +\delta\tau \, \pl_\tau \delta \Gamma_j(\dtss)
  +\half \delta\tau ^2 \: \pl^2_\tau \Gamma\sij(\dtss)
  .
\end{align} 
To find the perturbed time shift that maximizes the new
cross-correlogram, we set its derivative with respect $\delta\tau$ to
zero, thus requiring
\begin{align}\label{eq:Gamma_derivative}
  \pl_{\delta\tau}\tilde{\Gamma}\sij (\dtss+\delta\tau) 
  &= \delta\tau \, \pl^2_\tau \Gamma\sij(\dtss)+
  \pl_\tau \delta \Gamma_i(\dtss) +
  \pl_\tau \delta \Gamma_j(\dtss) =0.
\end{align} 
The solution then yields the cross-correlation traveltime perturbation
$\delta\dtss$ due to the model perturbation, namely
\begin{align}
  \label{dtss}
  \delta\dtss
  &=-\fracd{\pl_\tau \delta \Gamma_i(\dtss) + \pl_\tau \delta \Gamma_j(\dtss)}
      {\pl^2_\tau \Gamma\sij(\dtss)} =  
      \frac{\intot \pl_t s_j(t-\dtss)\,\delta s_i(t)\dt - \intot \pl_t s_i(t+\dtss)\,\delta s_j(t)\dt} 
           {\intot \pl^2_t s_i(t+\dtss) s_j(t)\dt}.
\end{align} 
We introduce a notation for the denominator,
\begin{equation}
  N\sij=\intot \pl^2_t s_i(t+\dtss) s_j(t)\dt,
\end{equation}
and rewrite eq.~(\ref{eq:derivative_misfit_CC_DD}), the derivative of
the differential cross-correlation objective function in
eq.~(\ref{eq:misfit_CC_DD}), as
\begin{align}\label{dchi}
  \delta \chi\CC\DD
  &= \intot  \left\{
  \sum_{i}  \left[ \sum_{j>i}   
    \ddtN  \pl_t s_j(t-\dtss) \right] \delta s_i(t)   
  -\sum_{j}  \left[  \sum_{i<j}
    \ddtN \pl_t s_i(t+\dtss)
    \right]  \delta s_j(t)  \right\}\dtt 
  .
\end{align}
Any scheme by which we minimize eq.~(\ref{eq:misfit_CC_DD}) proceeds
until model perturbations no longer produce meaningful adjustments
$\delta\dtss$ in the differential measurement on the synthetics
$\dtss$, in which case $\delta \chi\CC\DD$ effectively vanishes and a
minimum $\chi\CC\DD$ is reached.

\subsection{Adjoint source}

The translation of the gradient expression~(\ref{dchi}) into an
algorithm for which the spectral-element forward-modeling software can
be used to produce linearized expressions of the kind in
eq.~(\ref{kp}) again draws upon the Born approximation and the
reciprocity of the Green functions of the wave equation
\cite[]{Tromp+2005}. An alternative derivation is through the lens of
wave-equation constrained functional minimization
\cite[]{Liu+2006}. Either route leads to a paired set of adjoint
sources, formulated with respect to the synthetic wavefields $s_i$ and
$s_j$ recorded at locations~$\bx_i$ and $\bx_j$. Considering all
possible such pairs $i$ and $j$, we write the adjoint sources
explicitly as
\begin{subequations}
  \label{eq:adj_CC_DD}  
  \begin{align}
    \label{eq:adj_CC_DD_1}  
    f_i^{\dagger}(\bx,t)
    &=+\sum_{j>i} \ddtN \, \pl_t s_j \big( T-[t-\dtss] \big)  
    \delta (\bx-\bx_i)
    ,\\
    \label{eq:adj_CC_DD_2}
    f_j^{\dagger}(\bx,t)
    &=-\sum_{i<j} \ddtN \, \pl_t s_i \big( T- [t+\dtss] \big) 
    \delta (\bx-\bx_j)
    .
  \end{align}
\end{subequations}
The comparison of the expression for the adjoint sources of the
classical case in eq.~(\ref{eq:adj_CC}) with its double-difference
counterpart in eqs~(\ref{eq:adj_CC_DD}) reveals that the former only
involves one waveform per station, whereas in the latter case the
adjoint source for each station comprises the waveforms of all the
other stations with which it is being compared. As in the classical
case, however, the final adjoint source for double-difference
measurements is the sum of the contributions from all stations, which
are thus back-propagated simultaneously to obtain the adjoint kernels
of the objective function in eqs~(\ref{eq:misfit_CC_DD}).  All the
pairs of differential measurements are included in the adjoint
process, which is otherwise identical to that used in conventional
adjoint tomography. Since we linearly combine adjoint sources rather
than combining individual adjoint kernels for each data ensemble, only
one adjoint simulation is required to numerically evaluate the
gradient of the double-difference misfit function, and once again we
obtain an expression of the type in eq.~(\ref{kp}), with the summation
over the parameter type~$m$ implied:
\begin{equation}\label{kp2}
  \delta\chi\CC\DD=\int_\oplus K\DD_m(\bx) \,m(\bx)\dx
  ,
  \with
  K\DD_m(\bx)~\mbox{the double-difference cross-correlation traveltime misfit sensitivity kernel for}~m
  .
\end{equation}
A positive $\dd t\sij$ defined in eq.~(\ref{eq:cc_DD}) results in
positive kernel values~(\ref{kp2}) along the path between the source
and station~$j$, and negative kernel values along the path towards
station~$i$.  Therefore, to decrease the double-difference misfit
function in eq.~(\ref{eq:misfit_CC_DD}), the model update moves in the
negative kernel direction, increasing the wavespeed to station~$i$,
reducing it towards station~$j$, and decreasing the advancement of
$s_j$ over $s_i$.

\subsection{Cluster analysis and regularization}

From $N$ individual measurements a total of ${{N}\choose{2}}=n(n-1)/2$
double-difference measurements can be created. To enhance the
stability of the inverse problem and reduce the computational cost of
the data pairings, we can regularize the problem via cluster analysis
in which only pairs that are well linked to each other are
included. In general this can be achieved by incorporating pairs of
stations that are relatively `close', which enhances accuracy and
improves tomographic resolution. The maximum separation, however,
depends on the wavelength $\lambda$ and the path length~$L$ of the
phases considered, the scale length of the
heterogeneity, and the overall geometry of the illumination of the two
stations by seismic sources. In practice a physical distance
proportional to $\sqrt{\lambda L}$, the width of the first Fresnel
zone, is recommended \cite[e.g.,][]{Woodward92,Baig+2003}.  We can
furthermore exclude any station pairs whose spatial offset is much
smaller than the expected scale lengths of the structural variations.

Another possible method of clustering is to select station pairs with
`similar' signals, as measured by the correlation
\begin{align}\label{eq:cc_coherency}
  r\sij&=
  \frac{\Lambda\sij(\dtso)}{\sqrt{\Lambda_{ii}(0) \Lambda_{jj}(0)}},
  \qquad
  -1\le r\sij \le 1
  ,
\end{align} 
where we recall the definition of $\Lambda_{ij}$ from
eq.~(\ref{eq:cc_dd}). Similarity pairs can be defined on the basis of
threshold values for the correlation \cite[]{VandeCar+90}, or the
squared correlation can be used as weights
\cite[]{Waldhauser+2000}. The similarity criterion can be used alone,
or in combination with the relative physical proximity of the
stations.

\section{T~H~E{\hsps}~A~D~V~A~N~T~A~G~E{\hsps}O~F{\hsps}D~I~F~F~E~R~E~N~T~I~A~L{\hsps}M~E~A~S~U~R~E~M~E~N~T~S}

In this section we briefly list some of the properties of differential
traveltime measurements that make them good candidates for being less
affected by source uncertainties than traditional absolute
measurements. In the next section, these properties will be verified
and illustrated by numerical experiments.

\subsection{Scaling invariance}

The cross-correlation measurement in eq.~(\ref{eq:cc_ss}) is invariant
to scaling of the seismograms $s_i(t)$ and $s_j(t)$ by a constant
factor~$\alpha$:
\begin{align}\label{eq:cc_scaling}
  \argmax_\tau \intot [\alpha s_i(t+\tau)][\alpha s_j(t)] \dt 
  =\argmax_\tau  \left[\alpha^2 \intot s_i(t+\tau)  s_j(t) \dt \right]
  = \argmax_\tau \intot s_i(t+\tau) s_j(t) \dt =\dtss.
\end{align}
The scale invariance directly removes any sensitivity to incorrect
seismic moment estimates.

\subsection{Shift invariance}

The time shift calculated by maximizing the cross-correlogram in
eq.~(\ref{eq:cc_ss}) is unchanged if time is globally shifted by some
constant~$\tau_0$:
\begin{align}\label{eq:cc_shift}
  \argmax_\tau \intot s_i(t+\tau-\tau_0) \: s_j(t-\tau_0) \dt
  = \argmax_\tau \intot s_i(t+\tau)\; s_j(t) \dt =\dtss.
\end{align} 
The shift invariance directly removes any sensitivity to errors in
origin time.

\subsection{Source wavelet invariance} 

Knowledge of the source-time-function or `wavelet' is of course
crucial to making accurate measurements. Differential measurements,
from the same seismic event to a pair of stations, are more robust to
errors in the source wavelet than absolute ones. They should also be
relatively robust to errors in the source's focal mechanism, unless
incorrect assumptions should place one of the event-station paths ---
but not both --- on the wrong side of the nodal plane. The
cross-correlation time delays between two synthetic seismograms
calculated with the wrong source information should be relatively
robust as long as we can view the `wrong' records as filtered, or
otherwise mostly linearly mapped, from what would be the `right'
seismograms. As long as the phases targeted by the cross-correlation
time window remain well separated temporally from neighboring phases,
in either case, they could be remapped to the same impulsive arrivals
after deconvolution of their respective input source wavelets, hence
preserving the differential delay time.

\subsection{Differential sensitivity versus sensitivity of differential measurement}

Eq.~(\ref{dtss}) contains two terms that contribute to the traveltime
sensitivity, the Fr\'echet derivatives for the differential
cross-correlation traveltime. Both of these mix contributions from
both seismograms $s_i$ and $s_j$. Now suppose waveforms of $s_i(t)$
and $s_j(t)$ are identical upon time shifting, i.e., they are of the
`identical pulse shape' type (perhaps achieved by pre-processing,
e.g., to equate the Maslov indices of \textit{P} and \textit{PP}, or
\textit{S} and \textit{SS}) as identified by \cite{Dahlen+2000}. In
that case
\begin{align}
  s_i(t) \approx s_j(t-\dtss) \also  s_j(t) \approx s_i(t+\dtss)
  ,
\end{align}
and in that case eq.~(\ref{dtss}) would reduce to a form that does not
mix station indices, and we learn by comparison with eq.~(\ref{dti})
that
\begin{align}
  \label{eq:derivative_s1_app}
  \delta\dtss
  =  
  \frac{\intot\pl_t s_i(t)\delta s_i(t)\dt}{\intot \pl^2_t s_i(t)s_i(t)\dt}
  -
  \frac{\intot\pl_t s_j(t)\delta s_j(t)\dt}{\intot \pl^2_t
    s_j(t)s_j(t)\dt}
  =
  \delta\Delta t_i
  -\delta\Delta t_j
  . 
\end{align}
In other words, in the case of `identical' waveforms, the Fr\'{e}chet
sensitivity kernel for their differential traveltime can be
approximated by taking the difference of the individual Fr\'{e}chet
sensitivity kernels for each of the absolute traveltimes, and the
classical formalism of adjoint tomography can be used to perform
double-difference adjoint tomography, albeit approximately. This is
the approach adopted by, among others, \cite{Hung+2004}.

Conditional statements of the kind that validated the applicability
of~eq.~(\ref{eq:derivative_s1_app}) for using differential phases
(either for different phases at the same station, or from the same
phase at different stations) were made by \cite{Dahlen+2000} (compare
our eq.~\ref{dtss} with their eq.~75). These assumptions are, however,
not generally valid. In contrast, our eqs~(\ref{eq:adj_CC_DD}) are
proper generalizations of the differential traveltime adjoint sources
of \cite{Tromp+2005} (their eq.~61) which do not suffer similar such
restrictions --- save for the general adequacy of the Born
approximation, which appears practically unavoidable though perhaps
not theoretically unassailable
\cite[]{Hudson+81,Panning+2009,Wu+2014b}.

As a further point of note, it is computationally more efficient and
algorithmically less complicated to evaluate the double-difference
adjoint kernel using eqs~(\ref{eq:adj_CC_DD}) than via
eqs~(\ref{eq:adj_CC}) and~(\ref{eq:derivative_s1_app}). For example,
for a total number of $N$~measurements, one only needs one adjoint
simulation to numerically calculate the gradient of the differential
misfit under ${{N}\choose{2}}$ possible combinations. In contrast, by
using the approximation in eq.~(\ref{eq:derivative_s1_app}), one needs
to compute the Fr\'{e}chet sensitivity kernel for each individual
measurement, which requires $N$ adjoint simulations in
total. Subsequently these sensitivity kernels need to be combined by
weighted with the corresponding double-difference traveltime
measurements, and summed up for all grouped kernels. In conclusion,
the misfit sensitivity for differential measurements for use in
double-difference adjoint tomography is best obtained, both
theoretically and in practice, via the paired adjoint sources of
eqs~(\ref{eq:adj_CC_DD}).

\section{N~U~M~E~R~I~C~A~L{\hsps}E~X~P~E~R~I~M~E~N~T~S} 

Three types of validation experiments are conducted here. We consider
input checkerboard patterns but also a realistic phase speed
distribution, for sensor configurations that proceed in complexity
from two stations to simple arrays, and then to realistic array
configurations.

\subsection{Checkerboard model tests, local networks}

\begin{figure*}
  \rotatebox{-90}{\includegraphics[width=\model]{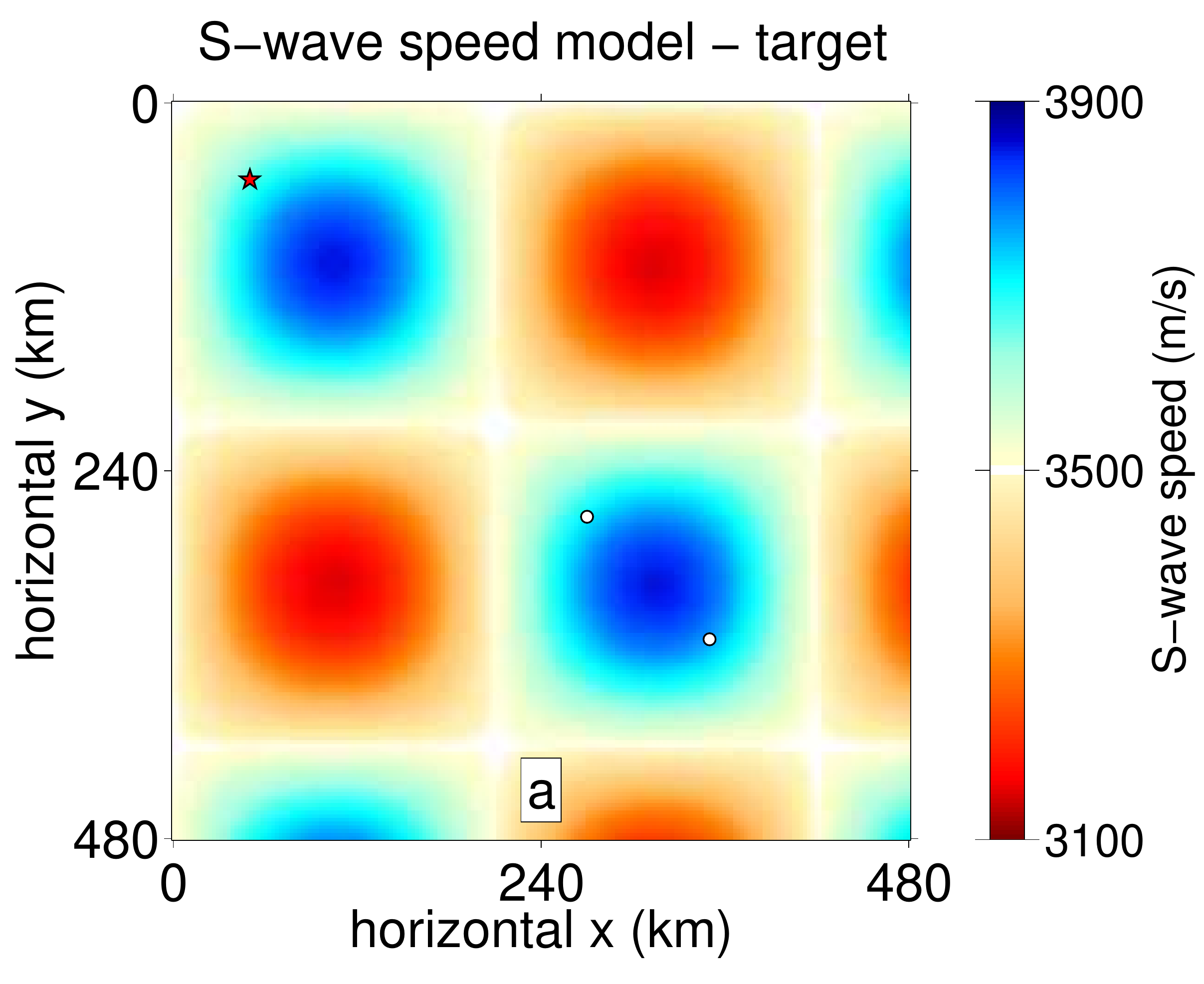}}\hspace{2em}
  \rotatebox{-90}{\includegraphics[width=\model]{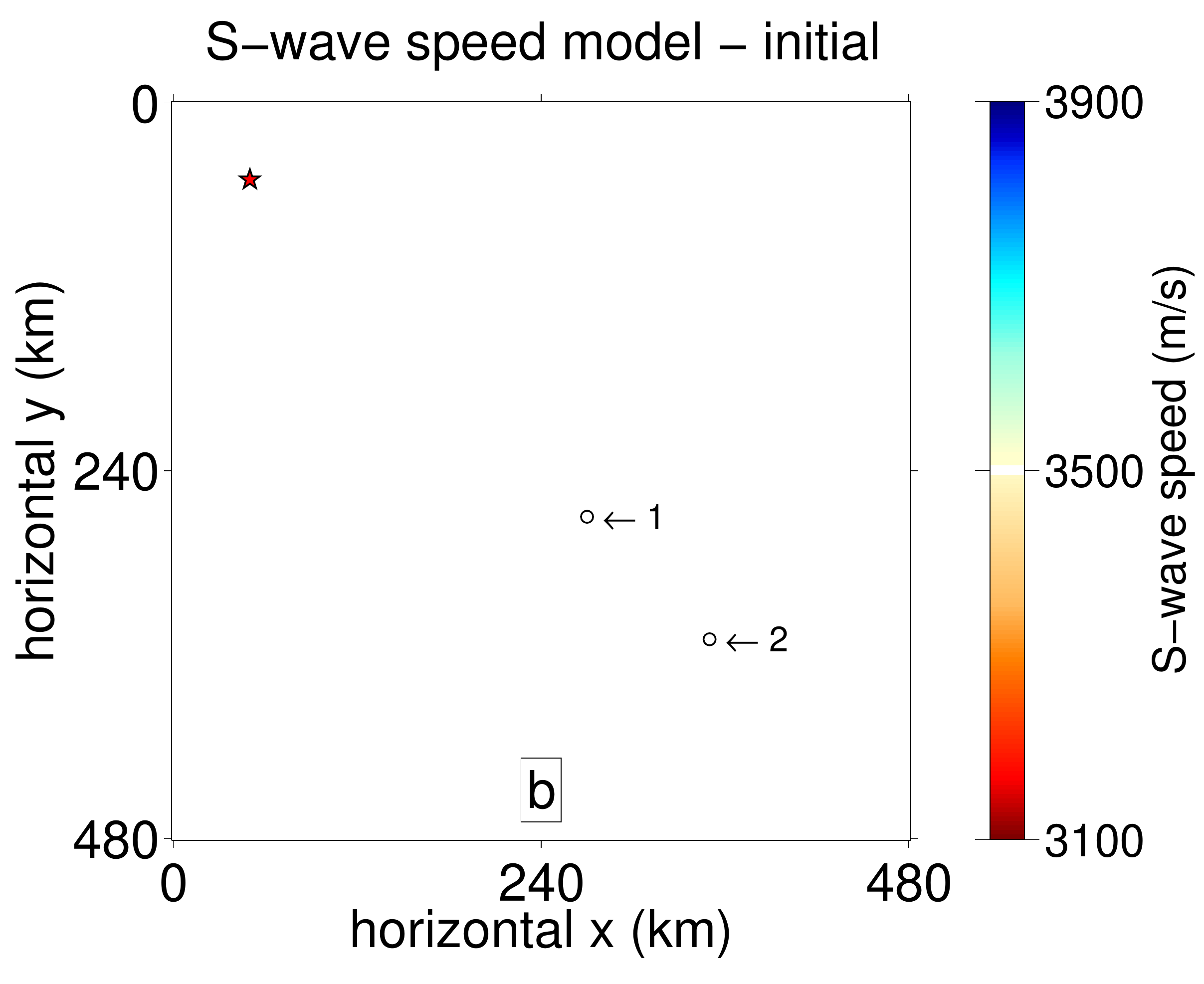}}
  \caption[Experiment~I: model configuration]{\label{fig:Tape_test1_model}
    Experiment~I: model configuration. Two stations at different
    epicentral distances located along a common source-receiver
    path. (a)~Target wavespeed model. (b)~Initial model. The source
    is depicted by a filled star. The open circles are stations $i=1$
    and $j=2$.} 
\end{figure*}

We first conduct a numerical experiment with strong lateral
heterogeneity, with a 2-D `checkerboard' input model as shown in
Fig.~\ref{fig:Tape_test1_model}. We use membrane surface waves
\cite[]{Tanimoto90,Peter+2007} as an analogue for short-period
(10--20~s) surface waves.  We use \texttt{SPECFEM2D}
\cite[]{Komatitsch+98} to model the wavefield. Absorbing boundary
conditions are employed on all sides.\enlargethispage{2em}

Experiment~I, as shown in Fig.~\ref{fig:Tape_test1_model}, involves
one source (filled star), and two stations (open circles labeled 1
and~2) that are aligned at different epicentral distances along a
common source-receiver ray path. The target model of interest is a
480~km $\times$ 480~km checkerboard expressed as four circular
\textit{S}-wave speed anomalies of alternating sign, shown in
Fig.~\ref{fig:Tape_test1_model}a. The initial model is a featureless
homogeneous model with \textit{S}-wave speed of 3500~m/s, shown in
Fig.~\ref{fig:Tape_test1_model}b. We uniformly discretize each
dimension using 40 elements of individual length 12~km. The source
wavelet used is the first derivative of a Gaussian function with a
dominant period of 12~s. The sampling rate is 0.06~s and the total
recording length is 4.8~s. The exact source-time function and event
origin time were used in the computation.

\begin{figure*}
  \includegraphics[angle=0,width=\traces]{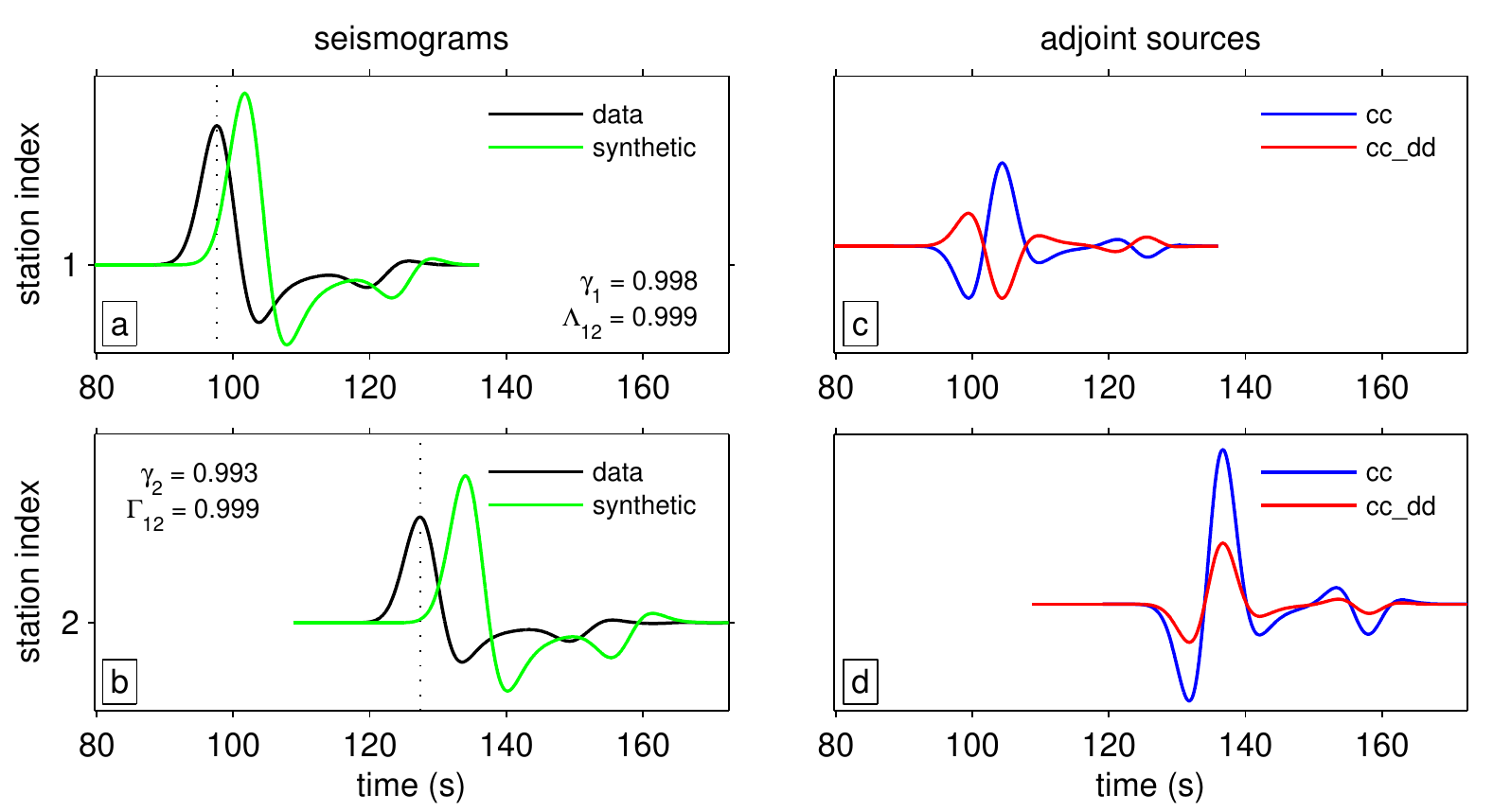}
  \caption[Experiment~I: data and synthetics]{\label{fig:Tape_test1_data}
    Experiment~I: data and synthetics. (a--b) Displacement
    seismograms, (a)~at station $i=1$ and (b)~at station $j=2$. The
    black traces are the `data', computed in the target model shown in
    Fig.~\ref{fig:Tape_test1_model}a.  The green traces are the
    `synthetics', calculated in the initial model shown in
    Fig.~\ref{fig:Tape_test1_model}b. Cross-correlation traveltimes,
    computed via eq.~(\ref{eq:cc_sd}), are $\Delta t_1=4.08$~s and
    $\Delta t_2=6.60$~s. The offset between the traces in panels (a)
    and~(b) is $\dtsoB=-29.76$~s for the observations and
    $\dtssA=-32.34$~s for the predictions, calculated via
    eq.~(\ref{eq:cc}). (c--d) Adjoint sources, (c)~at station $i=1$
    and (d)~at station $j=2$. The blue traces marked `cc' are
    calculated via eq.~(\ref{eq:adj_CC}) and correspond to the
    conventional cross-correlation metrics $\Delta t_1$ and $\Delta
    t_2$, rendered to scale.  The red traces marked `cc\_dd' are
    calculated via eq.~(\ref{eq:adj_CC_DD}) and correspond to the
    double-difference cross-correlation metric $\dtssA-\dtsoB=\dd
    t_{12}=-2.58$~s, to scale.}
\end{figure*}

Fig.~\ref{fig:Tape_test1_data}a and Fig.~\ref{fig:Tape_test1_data}b
show the displacement seismograms recorded at stations 1 and 2,
respectively. The black traces marked `data' are calculated in the
checkerboard model of Fig.~\ref{fig:Tape_test1_model}a. The green
traces labeled `synthetic' are predictions in the homogeneous model of
Fig.~\ref{fig:Tape_test1_model}b. We evaluate the time shift between
the predictions (green) and the observations (black) at each of the
stations via the conventional cross-correlation approach, as in
eq.~(\ref{eq:cc_sd}). These measurements are $\Delta t_1=4.08$~s
($d_1$ is advanced 4.08~s relative to $s_1$) and $\Delta t_2=6.60$~s
($d_2$ is advanced 6.60~s relative to $s_2$).  Following
eq.~(\ref{eq:misfit_CC}) in this example, the objective function is
the sum of the squares of the individual time shifts $\chi\CC=\half
([\Delta t_1]^2+[\Delta
  t_2]^2)=30.1$~s$^2$. Fig.~\ref{fig:Tape_test1_data}c and
Fig.~\ref{fig:Tape_test1_data}d show the adjoint sources calculated
from eq.~(\ref{eq:adj_CC}), in non time-reversed coordinates, as blue
lines labeled `cc'.

The double-difference cross-correlation approach measures, via
eq~(\ref{eq:cc_ss}), the time shift between the predictions made at
station~1 and~2 (green traces in Figs~\ref{fig:Tape_test1_data}a--b)
and, via eq.~(\ref{eq:cc_dd}), between the observations at those same
stations (black lines in Figs~\ref{fig:Tape_test1_data}a--b). The
first of these is $\dtssA=-32.34$~s ($s_2$ is delayed 32.34~s relative
to $s_1$), and the second $\dtsoB=-29.76$~s ($d_2$ is delayed 29.76~s
relative to $d_1$). Their `double' difference, from
eq.~(\ref{eq:cc_DD}), yields $\dd t_{12}=-2.58$~s (the delay of $s_2$
over $s_1$ is 2.58~s larger than that of $d_2$ over $d_1$).  The
double-difference objective function of eq.~(\ref{eq:cc_DD}) is the
sum of the squared double differences over all available pairs, which
in this case simply evaluates to $\chi\CC\DD=\half [\dd t_{12}]^2=
3.33$~s$^2$. Fig.~\ref{fig:Tape_test1_data}c and
Fig.~\ref{fig:Tape_test1_data}d show the adjoint sources calculated
from eqs~(\ref{eq:adj_CC_DD_1}) and~(\ref{eq:adj_CC_DD_2}), again in
natural time coordinates, as red lines, which are labeled `cc\_dd'.

\begin{figure*}
  \includegraphics[angle=-90,width=\kernels]{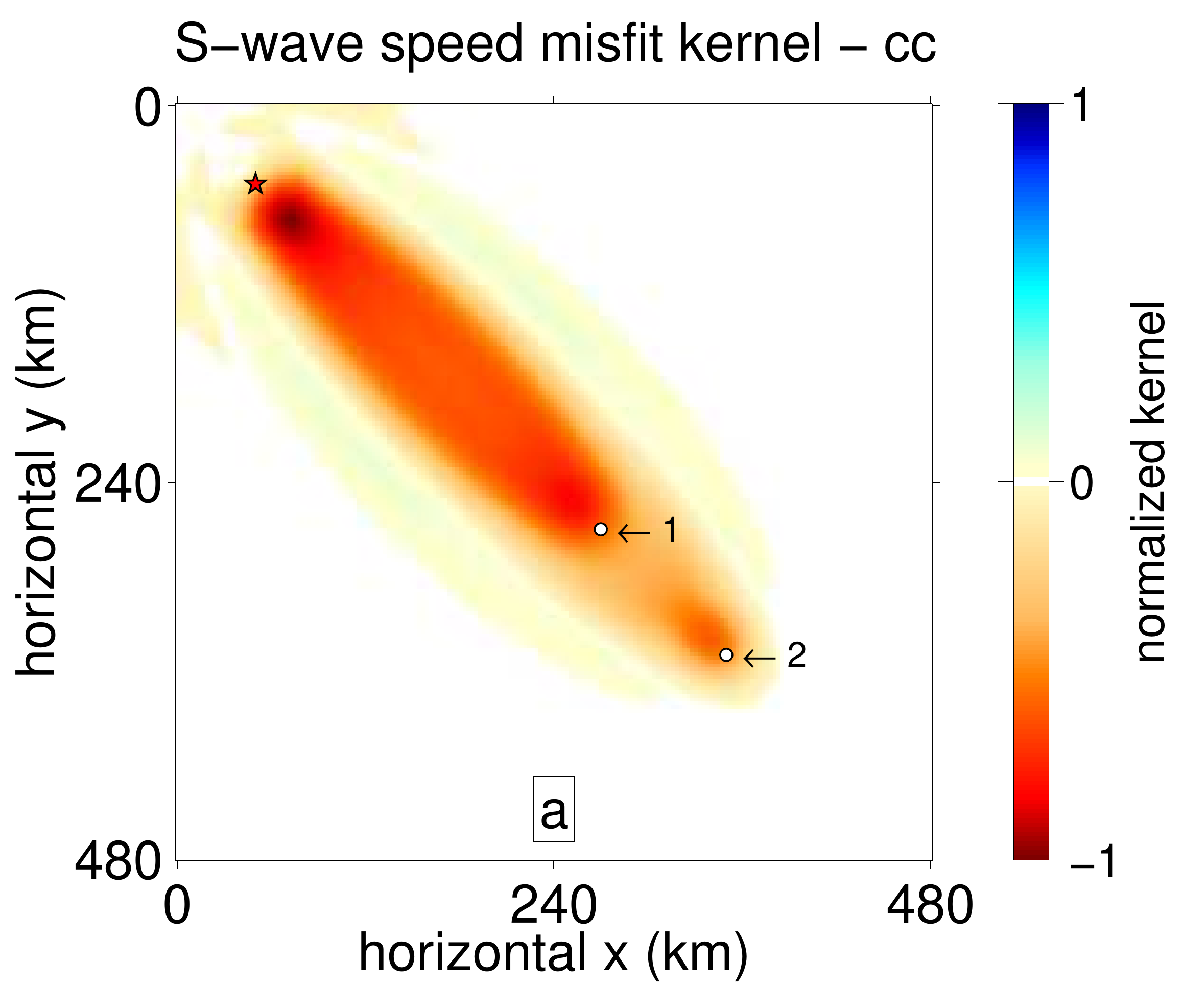}\hspace{2em}
  \includegraphics[angle=-90,width=\kernels]{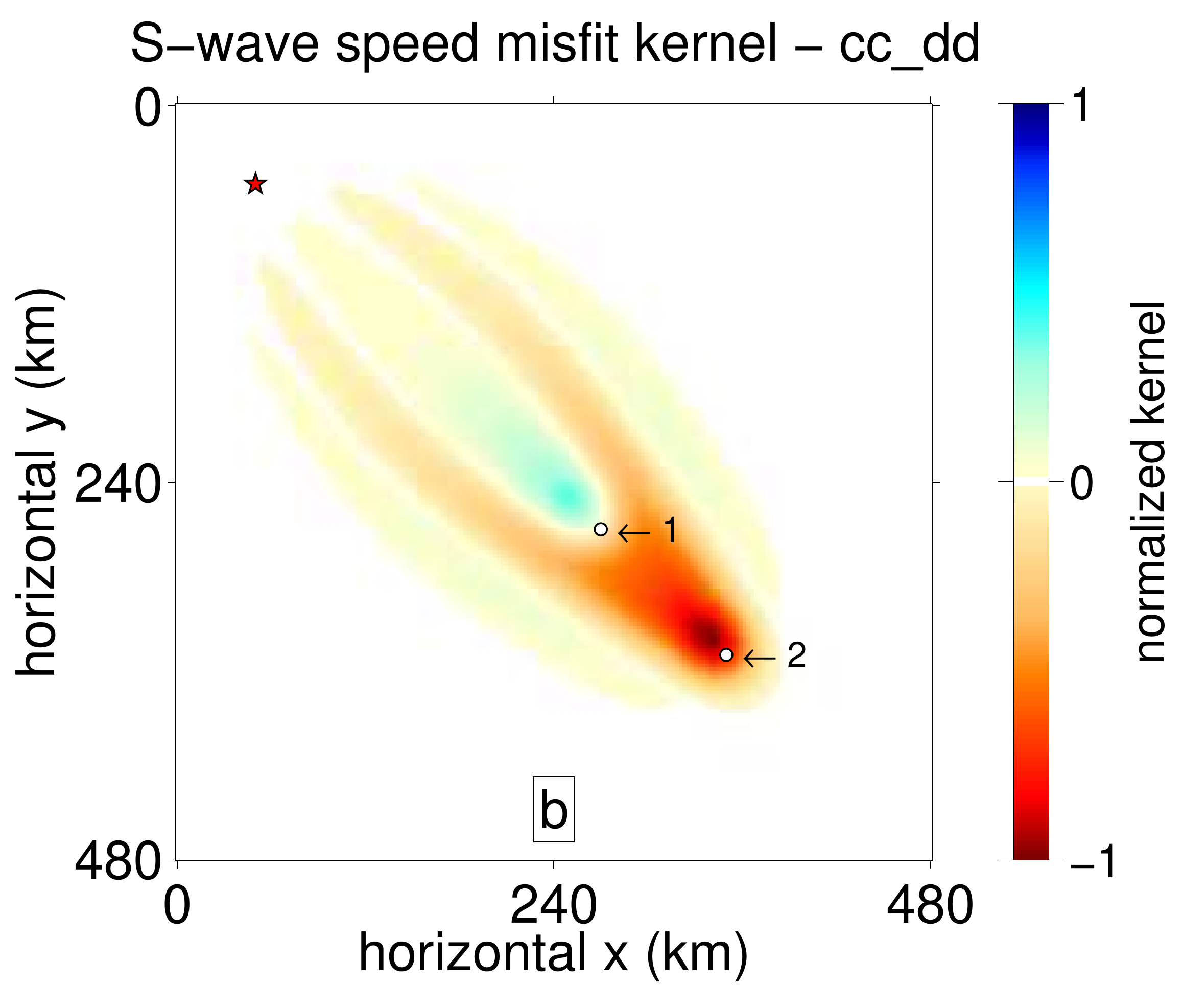}
  \caption[Experiment~I: misfit sensitivity kernels]{\label{fig:Tape_test1_kernel}
    Experiment~I: misfit sensitivity kernels. (a)~Cross-correlation
    traveltime misfit sensitivity kernel~$K_m(\bx)$, as in
    eq.~(\ref{kp}), computed using the adjoint sources shown by the
    blue lines in
    Fig.~\ref{fig:Tape_test1_data}c--d. (b)~Double-difference
    cross-correlation traveltime misfit sensitivity
    kernel~$K_m\DD(\bx)$, as in eq.~(\ref{kp2}) computed using the
    adjoint sources shown by the red lines in
    Fig.~\ref{fig:Tape_test1_data}c--d. In both cases, the parameter
    of interest is $m=\beta$, the \textit{S}-wave speed.}
\end{figure*}

Fig.~\ref{fig:Tape_test1_kernel} shows the misfit sensitivity kernels
for both types of measurements. The cross-correlation traveltime
\textit{S}-wave speed adjoint misfit kernel $K_\beta(\bx)$ shown in
Fig.~\ref{fig:Tape_test1_kernel}a is most sensitive to the area of
effective overlap of the ray paths from the source to either station.
This behavior is indeed explained by the nature of the kernel, which
is obtained by summation of the individual adjoint kernels for each
station. The desired model update is in the opposite kernel direction,
increasing the wavespeed to station~1 and to station~2.  In contrast,
the double-difference cross-correlation traveltime \textit{S}-wave
speed adjoint misfit kernel $K\DD_\beta(\bx)$ shown in
Fig.~\ref{fig:Tape_test1_kernel}b displays a dominant sensitivity in
the area in-between the two stations. The `double' differencing of the
traveltime metrics between the distinct stations, cancels out much of
the sensitivity common to both ray paths, the area between the source
and station~1 in this case. The traveltime delay of the waves recorded
at station~2, relative to station~1, mainly reflects the average
\textit{S}-wave speed between stations~1 and~2. The finite-frequency
character of the measurements remains apparent from the `swallowtails'
visible in the area between the source and station~1. The opposite
kernel direction required for model update is to slow down along the
path from the source to station~1, and to speed up towards
station~2. Since the path of station~1 overlaps that of station~2, the
resulting effect is to increase the wavespeed between station~1 and
station~2.

\begin{figure*}
  \rotatebox{-90}{\includegraphics[width=\model]{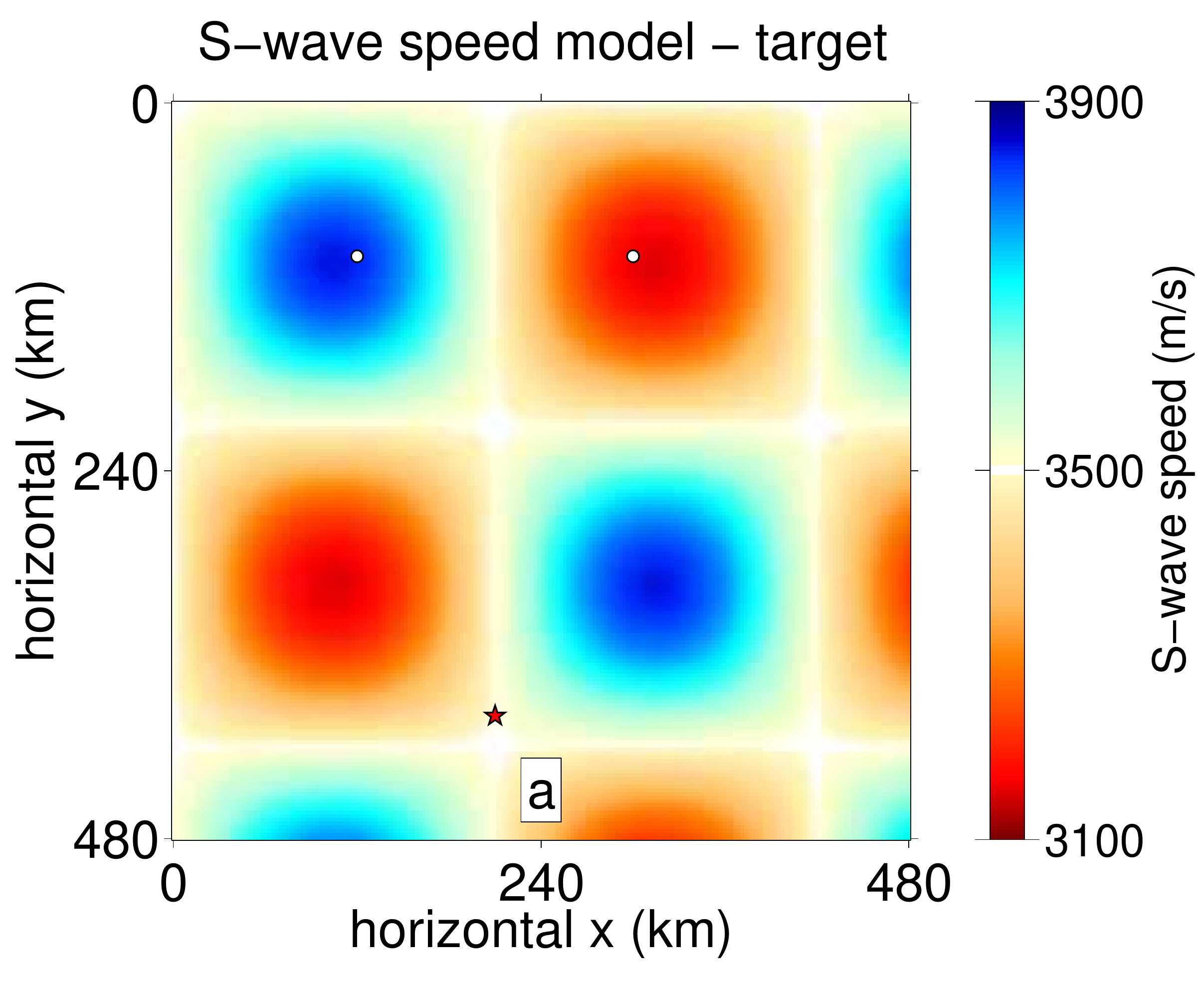}}\hspace{2em}
  \rotatebox{-90}{\includegraphics[width=\model]{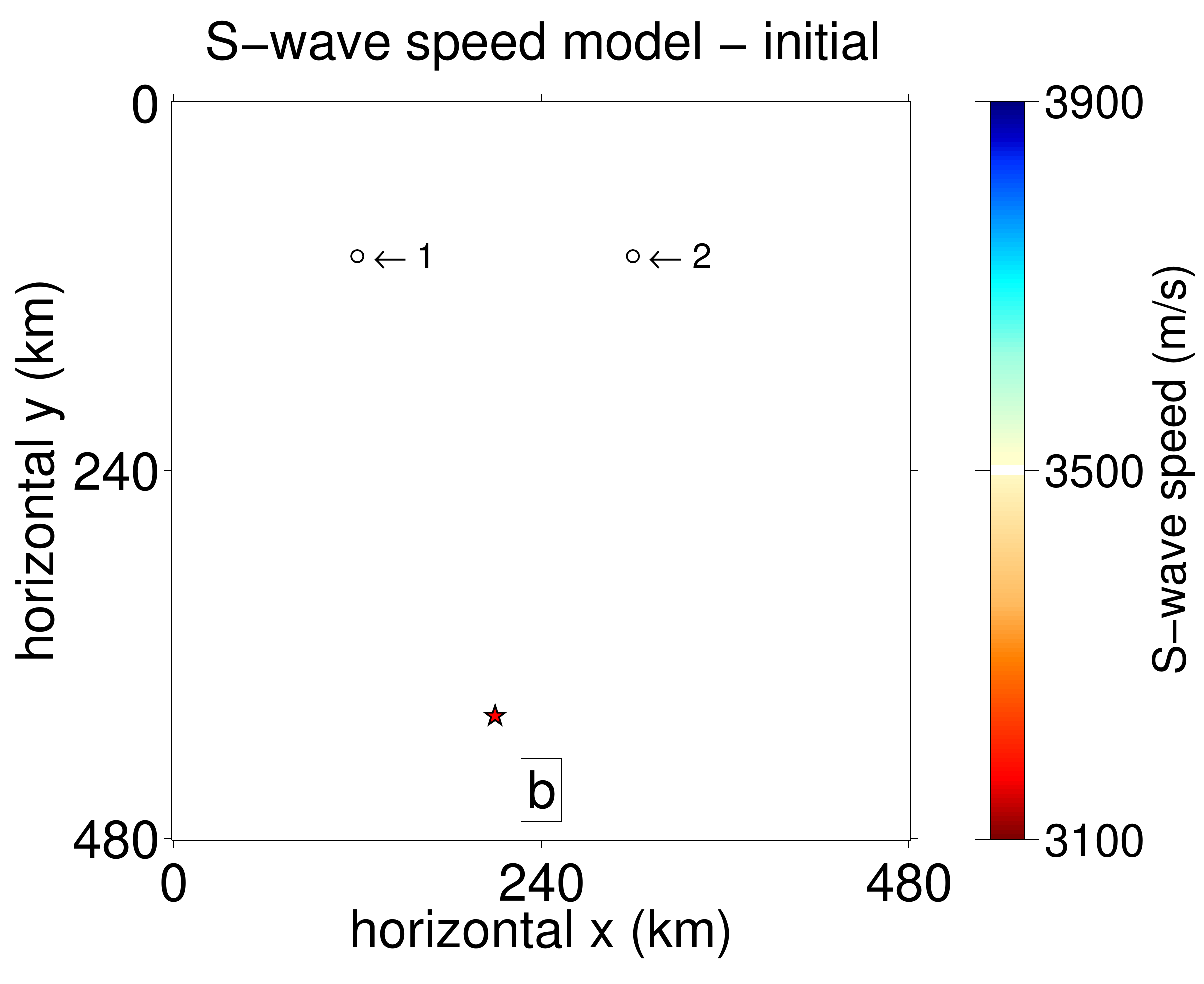}}
  \caption[Experiment~II: model configuration]{\label{fig:Tape_group2_model} 
    Experiment~II: model configuration. Two stations (open circles) at
    the same epicentral distance but with different azimuthal
    directions with respect to the source (filled star). (a)~Target
    wavespeed model. (b)~Initial model.}
\end{figure*}

Experiment~II, as shown in Fig.~\ref{fig:Tape_group2_model}, again
involves one source (filled star) and two stations (open circles), but
now the latter have been arrange to lie at the same epicentral
distance but at different azimuthal directions relative to the source.

\begin{figure*}
  \includegraphics[angle=0,width=\traces]{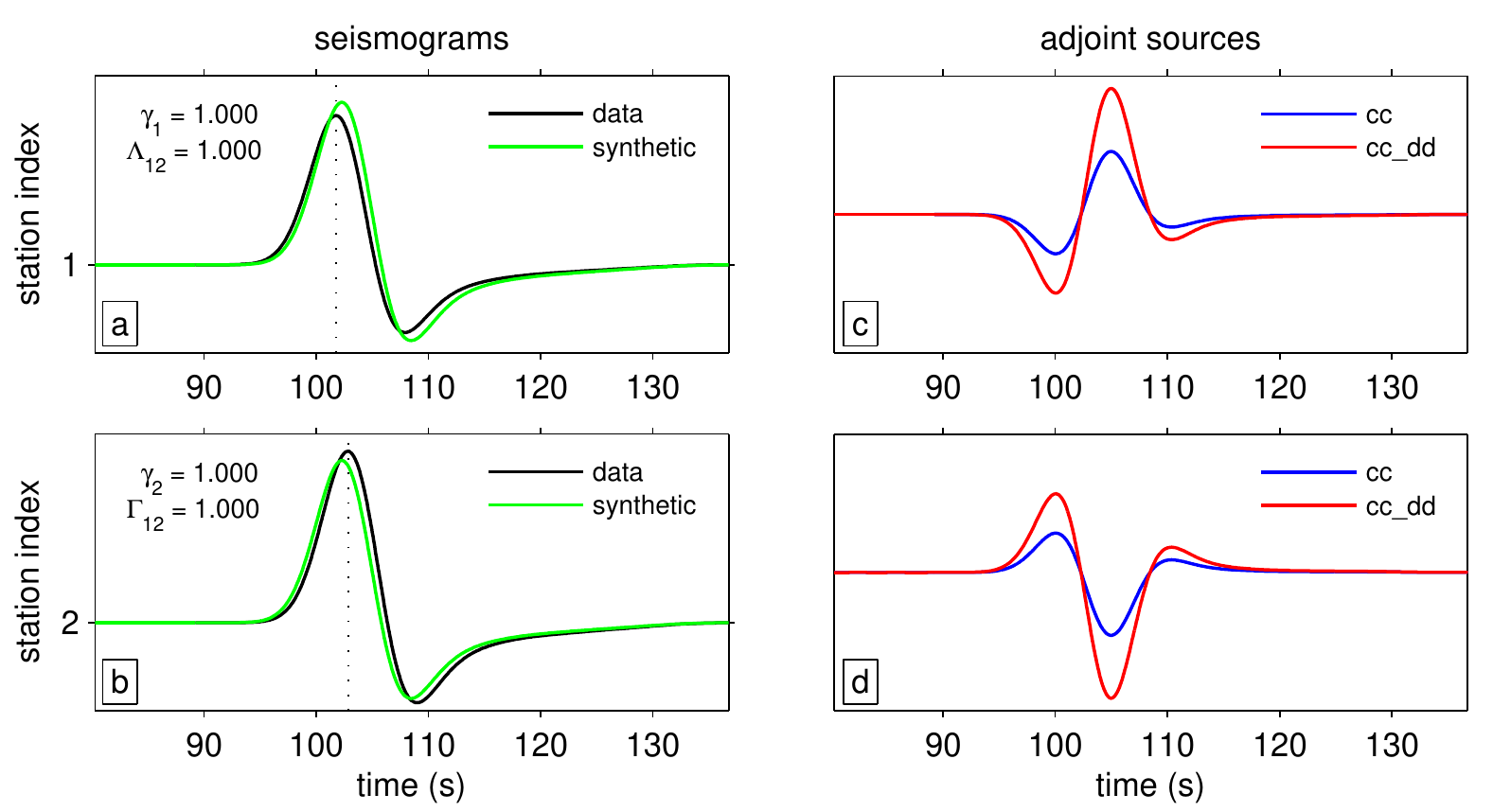}
  \caption[Experiment~IIA: data and synthetics]{\label{fig:Tape_test4_data} 
    Experiment~IIA: data and synthetics. (a--b) Displacement
    seismograms at stations~1 and~2. Black traces are computed in the
    target model of Fig.~\ref{fig:Tape_group2_model}a. Green traces
    are for the initial model in
    Fig.~\ref{fig:Tape_group2_model}b. The correct source-time
    function and origin time are being used. Cross-correlation
    traveltimes $\Delta t_1=0.54$~s and $\Delta t_2=-0.54$~s.
    Double-difference measurements $\dtsoB=-1.08$~s, $\dtssA=0.00$~s,
    hence $\dd t_{12}=1.08$~s. (c--d) Adjoint sources.  Blue traces
    are for the conventional cross-correlation (`cc') adjoint
    source~(\ref{eq:adj_CC}). Red traces are for the
    double-difference cross-correlation (`cc\_dd') adjoint
    source~(\ref{eq:adj_CC_DD}).}
\end{figure*}

\begin{figure*}
  \includegraphics[angle=-90,width=\kernels]{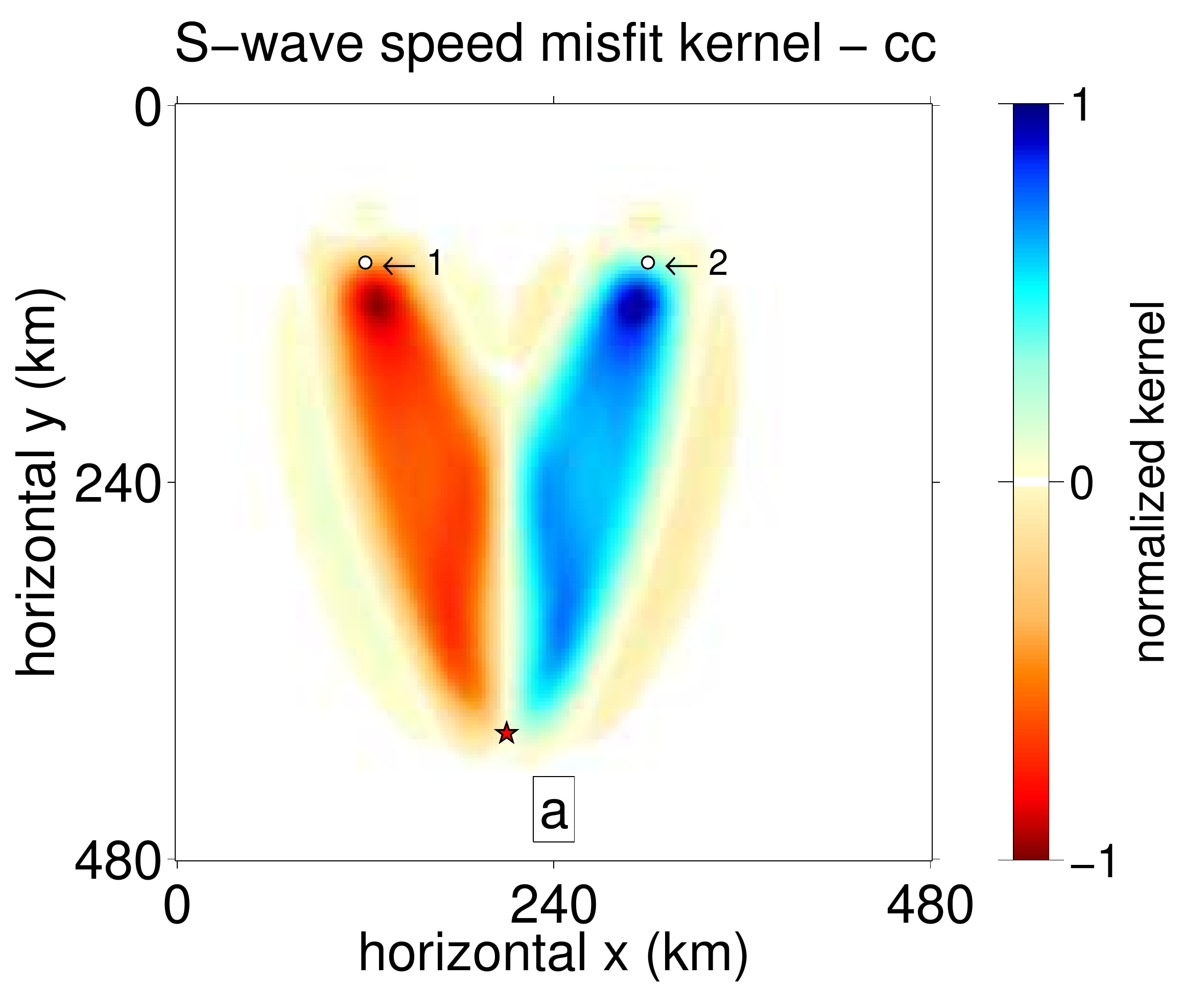}\hspace{2em}
  \includegraphics[angle=-90,width=\kernels]{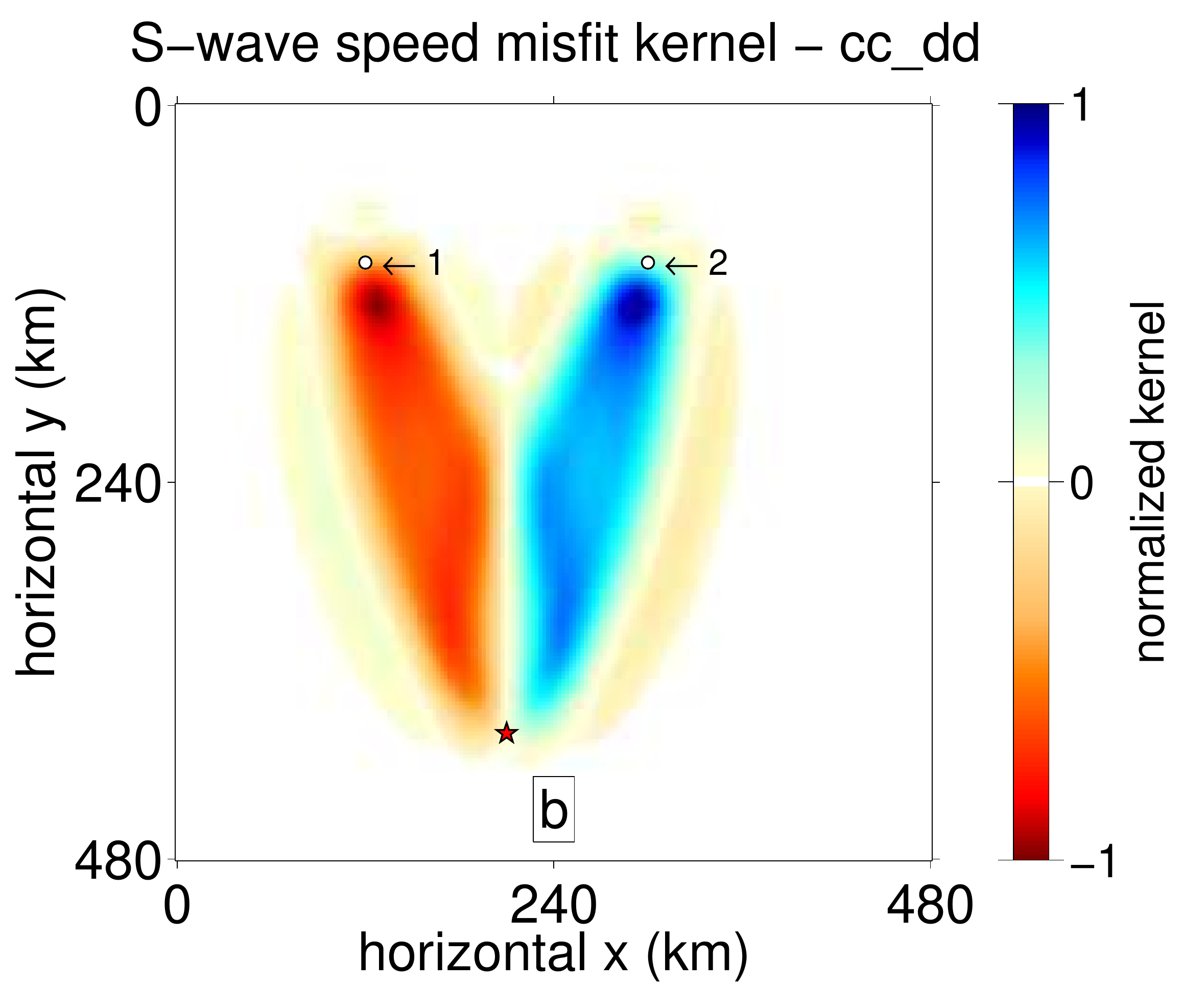}
  \caption[Experiment~IIA: misfit sensitivity kernels]{\label{fig:Tape_test4_kernel}
    Experiment~IIA: misfit sensitivity kernels.  (a)~Adjoint
    kernel~$K_\beta(\bx)$ corresponding to the adjoint sources (blue
    lines) in
    Fig.~\ref{fig:Tape_test4_data}c--d. (b)~Double-difference adjoint
    kernel~$K_\beta\DD(\bx)$ from the adjoint sources (red lines)
    in Fig.~\ref{fig:Tape_test4_data}c--d. The kernels look very
    similar but their role and interpretation are different.}
\end{figure*}

In Experiment~IIA we consider the ideal situation when source
information is known. The exact timing and the exact source wavelet
(again, the first derivative of a Gaussian with dominant period of
12~s) are used for synthetic modeling. Fig.~\ref{fig:Tape_test4_data}a
and Fig.~\ref{fig:Tape_test4_data}b show the displacement seismograms
recorded at both stations, with the checkerboard-model traces in black
and the homogeneous-model traces in green. The cross-correlation time
shift measurements are now $\Delta t_1=0.54$~s ($d_1$ is advanced
0.54~s relative to $s_1$) and and $\Delta t_2=-0.54$~s ($d_2$ is
delayed 0.54~s relative to $s_2$).  These are small and almost
identical as the ray paths sample almost equal path lengths of slow
and fast wavespeed anomalies. Fig.~\ref{fig:Tape_test4_data}c and
Fig.~\ref{fig:Tape_test4_data}d show the adjoint sources for this
model setup. In blue, for the traditional cross-correlation time shift
between predictions and observations, in red for the difference of the
differential cross-correlation time shifts between stations, where
$\dtssA=0.00$~s (as $s_1$ and $s_2$ arrive at the same time),
$\dtsoB=-1.08$~s (as $d_2$ is delayed by 1.08~s relative to $d_1$), and
the `double' difference $\dd
t_{12}=1.08$~s. Fig.~\ref{fig:Tape_test4_kernel} shows the misfit
sensitivity kernels for both types of adjoint measurements.



\begin{figure*}
  \includegraphics[angle=0,width=\traces]{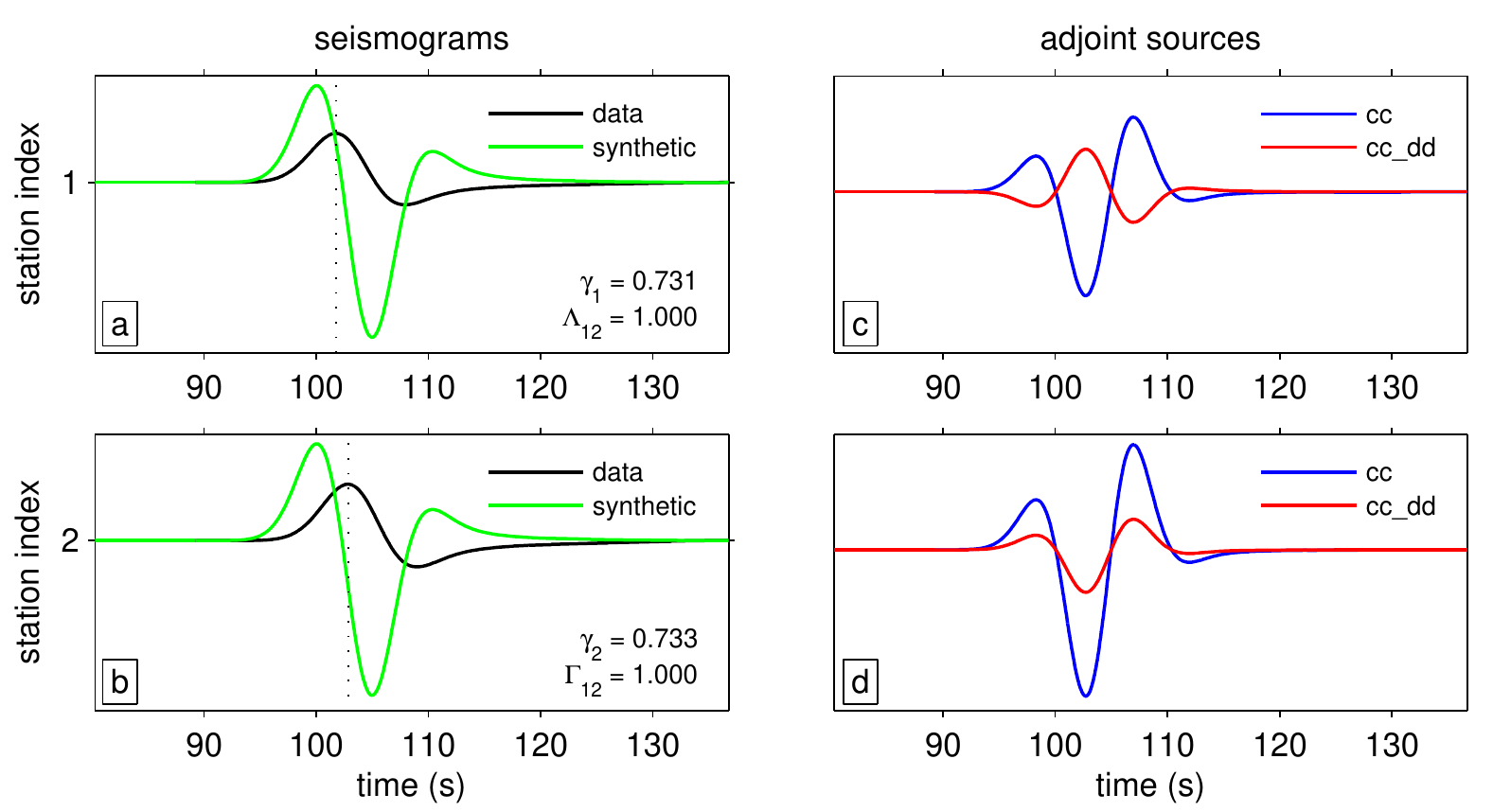}
  \caption[Experiment~IIB: data and synthetics]{\label{fig:Tape_test5_data} 
    Experiment~IIB: data and synthetics. (a--b) Displacement
    seismograms, computed in the target model (black) and in the
    initial model (green). An incorrect source model was used in the
    latter case (different source-time function but the same origin
    time and dominant periods). Cross-correlation traveltimes $\Delta
    t_1=-2.64$~s and $\Delta t_2=-3.72$~s.  Double-difference
    measurements $\dtsoB=-1.08$~s, $\dtssA=0.00$~s, thus $\dd
    t_{12}=1.08$~s. (c--d) Adjoint sources for the conventional
    cross-correlation (blue) and for the double-difference
    cross-correlation (red).}
\end{figure*}
\begin{figure*}
  \includegraphics[angle=-90,width=\kernels]{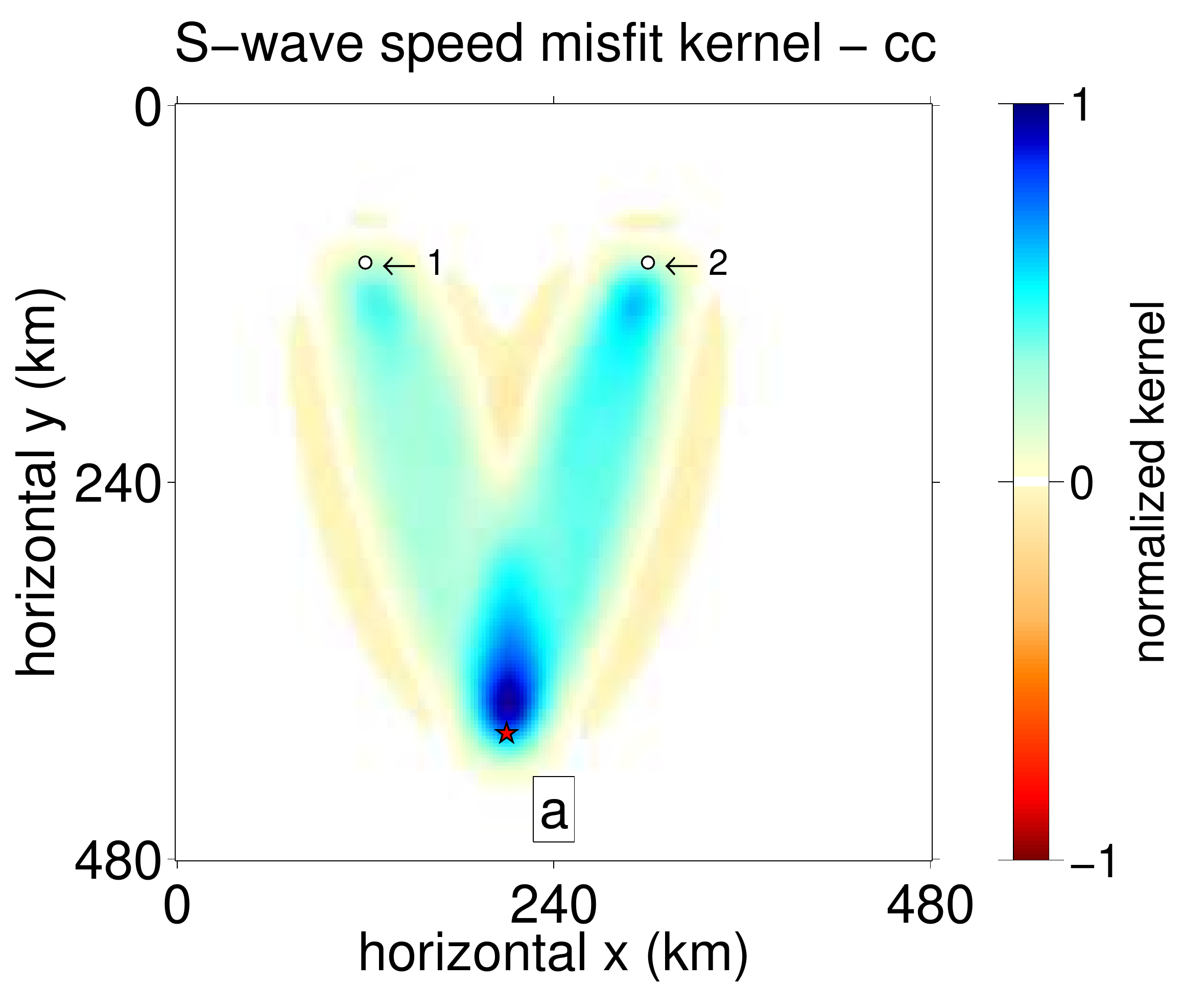}\hspace{2em}
  \includegraphics[angle=-90,width=\kernels]{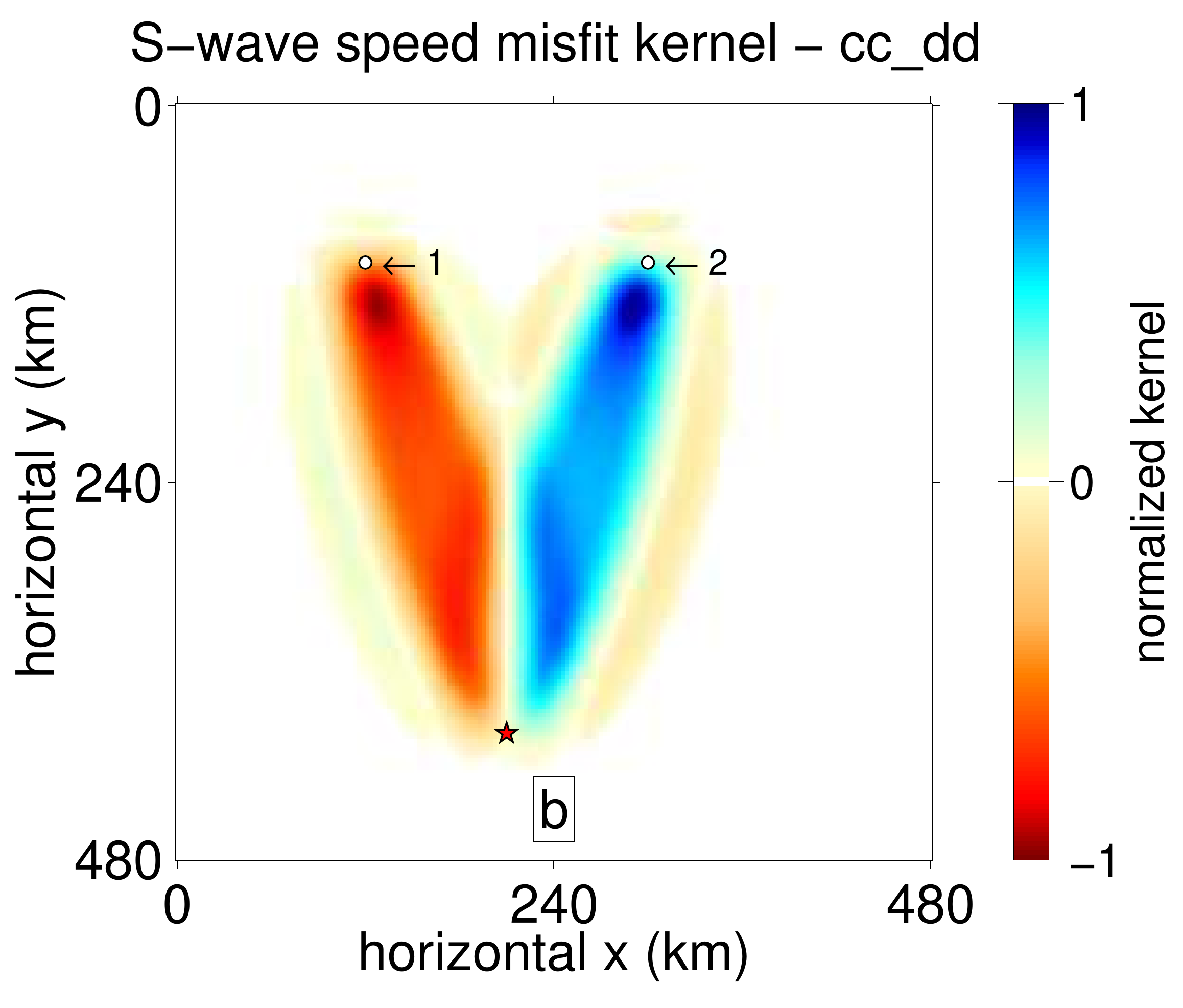}
  \caption[Experiment~IIB: misfit sensitivity kernels]{\label{fig:Tape_test5_kernel} 
    Experiment~IIB: misfit sensitivity kernels. (a)~Conventional
    adjoint kernel~$K_\beta(\bx)$ and (b)~double-difference adjoint
    kernel~$K_\beta\DD(\bx)$ from the adjoint sources shown in
    Fig.~\ref{fig:Tape_test5_data}c--d.  The kernels are very
    different from one another. Since the wrong source-time function
    was used, the cross-correlation traveltime misfit kernel in (a)
    differs greatly from the one in Fig.~\ref{fig:Tape_test4_kernel}a,
    whereas the double-difference misfit kernel in (b) is almost
    identical to the one shown in Fig.~\ref{fig:Tape_test4_kernel}b.}
\end{figure*}

The $K_m(\bx)$ and $K_m\DD(\bx)$ kernels, at first glance, look very
similar, but their physical meanings are very different. The
traditional measurements provide absolute information on wavespeed
structure along the ray paths, whereas the double-difference
measurements are sensitive to relative structural variations.  From
the classical prediction-observation cross-correlation measurements,
we learn that the surface waves predicted to be recorded at station~1
arrive 0.54~s later than the observations, and those predicted at
station~2 arrive 0.54~s earlier than the observations. To reduce the
misfit function (in eq.~\ref{eq:misfit_CC}), the wavespeed along the
ray path from the source to station~1 should be increased, and the
wavespeed between the source and station~2 decreased. On the other
hand, from the differential cross-station traveltime measurements, we
deduce that there is a 1.08~s advance in the arrival of the observed
waves at station~1 relative to station~2, but the synthetic records
arrive at exactly the same time at both stations, based on the lag
time of their cross-correlation maximum. Of course here too, to reduce
the misfit (in eq.~\ref{eq:misfit_CC_DD}), the wavespeed from the
common source to station~1 should be made faster than that to
station~2.

In Experiment~IIB we treat the more realistic scenario where an
incorrect source wavelet is used for the synthetic
modeling. Fig.~\ref{fig:Tape_test5_data}a
and~\ref{fig:Tape_test5_data}b show the displacement synthetics. The
observations (black) are in the checkerboard model and use the first
derivative of the Gaussian function for a source wavelet. The
predictions (green) use the second derivative of the Gaussian, the
`Ricker' wavelet. Both source wavelets share the same excitation time
and dominant period of 12~s. The cross-correlation time shifts between
the synthetics and observations are $\Delta t_1=-2.64$~s and $\Delta
t_2=-3.72$~s. These are markedly different from the measurements
obtained above using the correct source wavelet (recall those were
$\Delta t_1=0.54$~s, $\Delta t_2=-0.54$~s). In contrast, the
double-difference time shift between station~1 and station~2 is
unchanged from that which uses the correct source wavelet, $\dd
t_{12}= 1.08$~s. Fig.~\ref {fig:Tape_test5_kernel} shows the misfit
kernels.

The effects of an incorrect source signature on the traditional
cross-correlation measurements can be clearly seen by comparing the
adjoint kernel in Fig.~\ref {fig:Tape_test5_kernel}a with that shown
in Fig.~\ref{fig:Tape_test4_kernel}a. These are markedly
different. Fortunately, the double-difference measurement approach
yields relatively consistent adjoint kernels, based on the comparison
of the kernel shown in in Fig.~\ref{fig:Tape_test5_kernel}b with the
one using exact source signature in
Fig.~\ref{fig:Tape_test4_kernel}b. The latter two are nearly
identical, which is indicative of desirable robustness in the
inversion.

\begin{figure*}
  \includegraphics[angle=0,width=\traces]{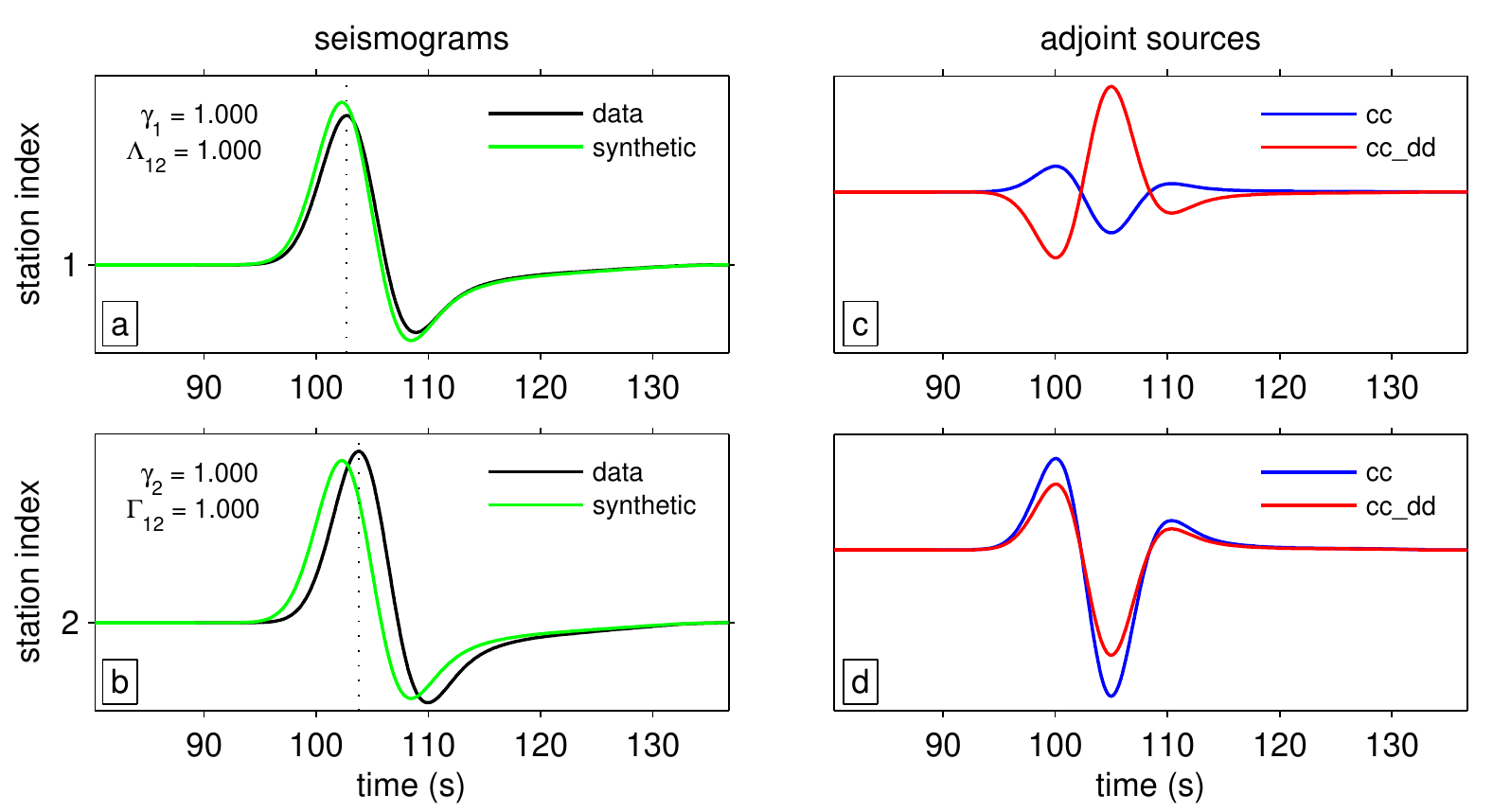}
  \caption[Experiment~IIC: data and synthetics]{\label{fig:Tape_test6_data} 
    Experiment~IIC: data and synthetics.  (a--b) Displacement records
    in the target (black) and in the initial model (green). In the
    latter case an incorrect source model had the same source-time
    function but a shifted origin time.  Cross-correlation traveltimes
    $\Delta t_1=-0.42$~s and $\Delta t_2=-1.50$~s.  Double-difference
    measurements $\dd t_{12}=1.08$~s. (c--d) Adjoint sources for the
    conventional (blue) and for the double-difference
    cross-correlation (red).}
\end{figure*}

\begin{figure*}
  \includegraphics[angle=-90,width=\kernels]{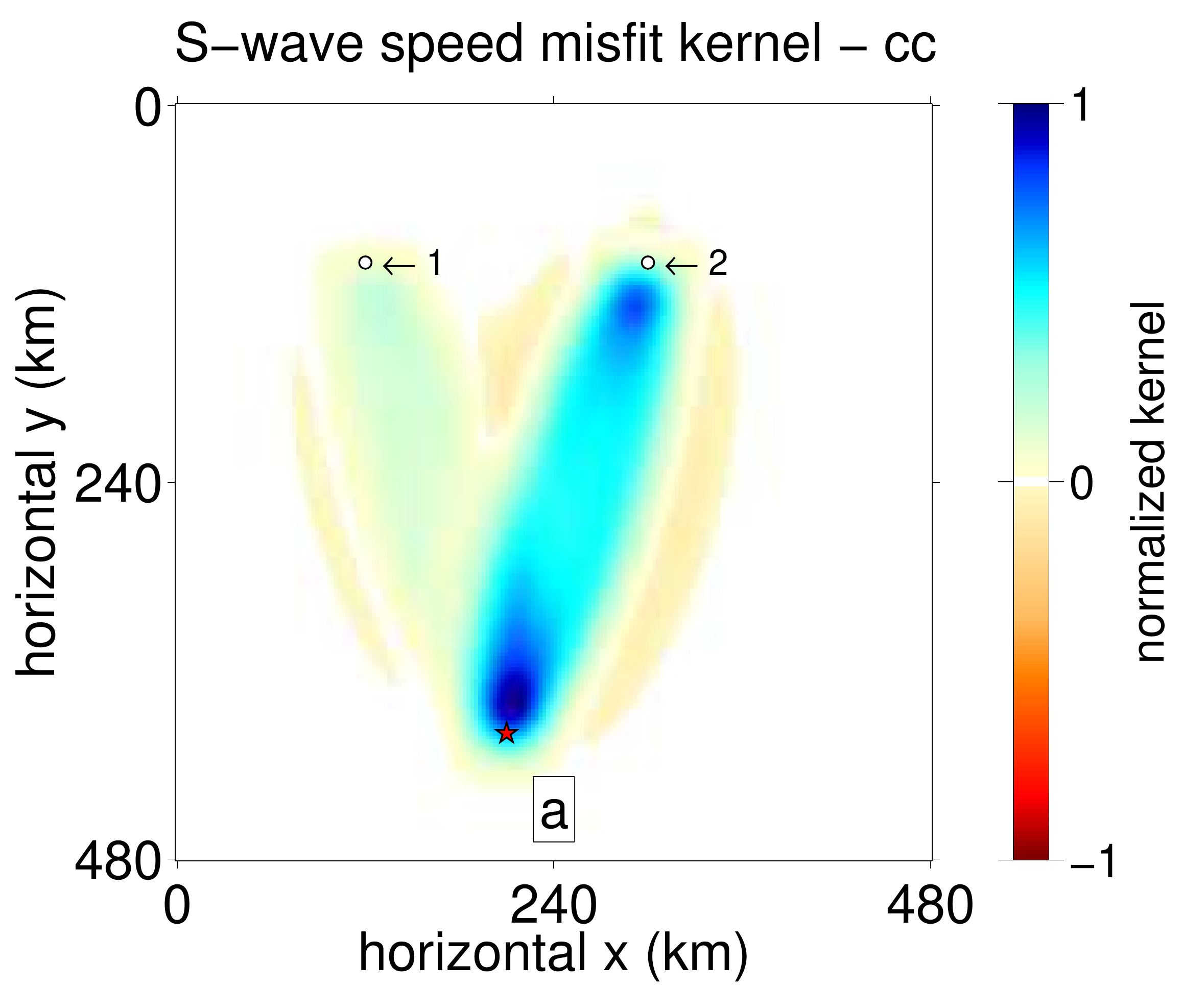}\hspace{2em}
  \includegraphics[angle=-90,width=\kernels]{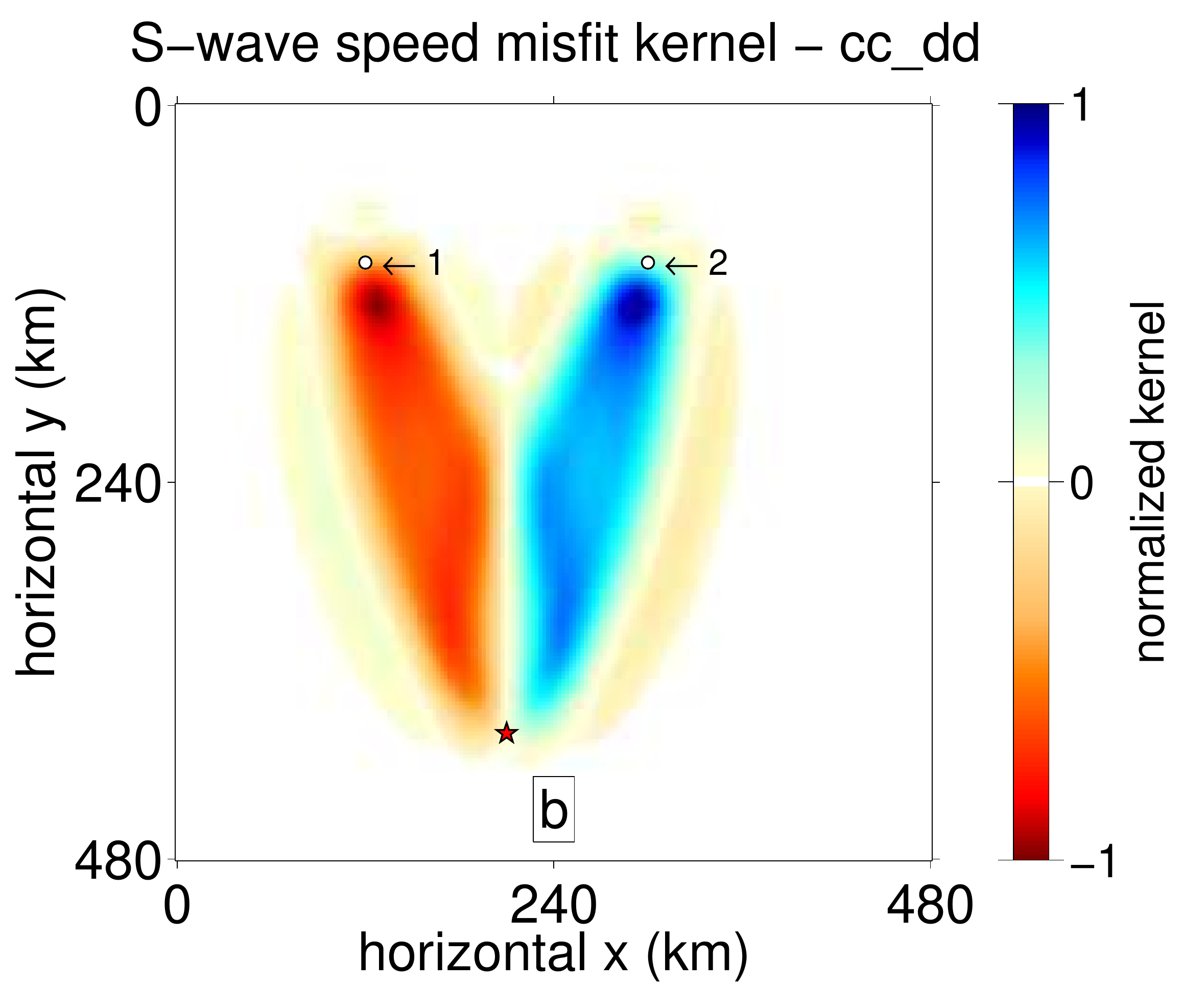}
  \caption[Experiment~IIC: misfit sensitivity kernels]{\label{fig:Tape_test6_kernel} 
    Experiment~IIC: misfit sensitivity kernels. (a)~Conventional
    kernel~$K_\beta(\bx)$ and (b)~double-difference
    kernel~$K_\beta\DD(\bx)$ from the adjoint sources  in
    Fig.~\ref{fig:Tape_test6_data}c--d.  The kernels are very
    different from one another, with (a) differing greatly from the
    one in Fig.~\ref{fig:Tape_test4_kernel}a, but (b) being almost
    identical to the one in Fig.~\ref{fig:Tape_test4_kernel}b.}
\end{figure*}

In Experiment~IIC we next consider the scenario of having an incorrect
origin time, an advance of 0.96~s expressed by the
synthetics. Fig.~\ref{fig:Tape_test6_data}a and
Fig.~\ref{fig:Tape_test6_data}b show the records, with the same color
conventions as before. It comes as no surprise that the
cross-correlation time shifts between predictions and observations are
biased. We obtain $\Delta t_1=-0.42$~s and $\Delta t_2=-1.50$~s, a
spurious shift of 0.96~s for each measurement. The differential
cross-correlation time shifts between stations, however, are not
affected at all, since the predictions at both stations, wrong as they
both are, are shifted in exactly the same way with respect to the true
source origin time. Fig.~\ref{fig:Tape_test6_kernel} shows the misfit
kernels. The conventional cross-correlation kernel, in
Fig.~\ref{fig:Tape_test6_kernel}a, are clearly affected by the timing
error, while the double-difference ones, in
Fig.~\ref{fig:Tape_test6_kernel}b, are not biased at all by any such
inaccuracies.

\begin{figure*}\centering
  \rotatebox{-90}{\includegraphics[width=\model]{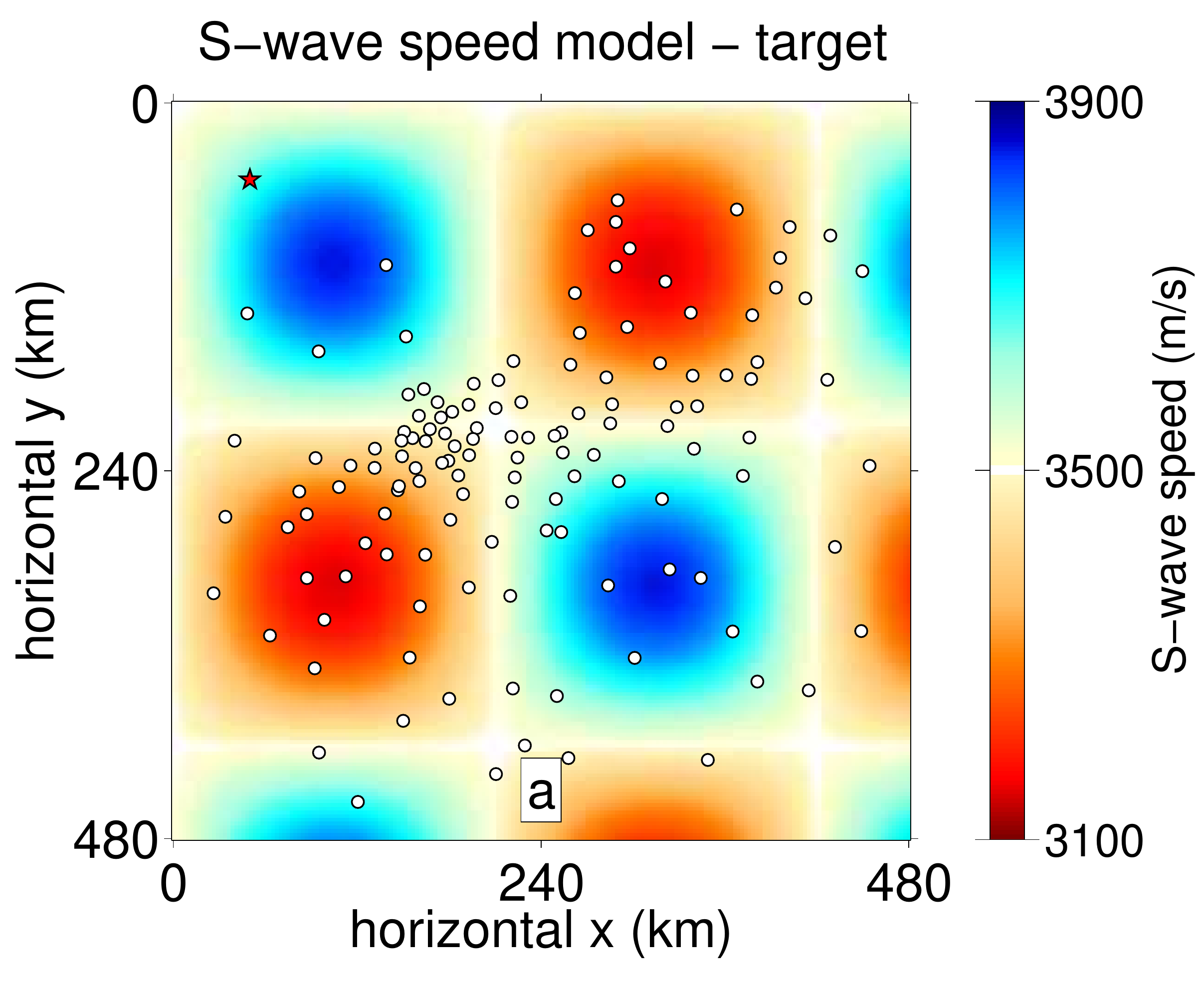}}\hspace{2em}
  \rotatebox{-90}{\includegraphics[width=\model]{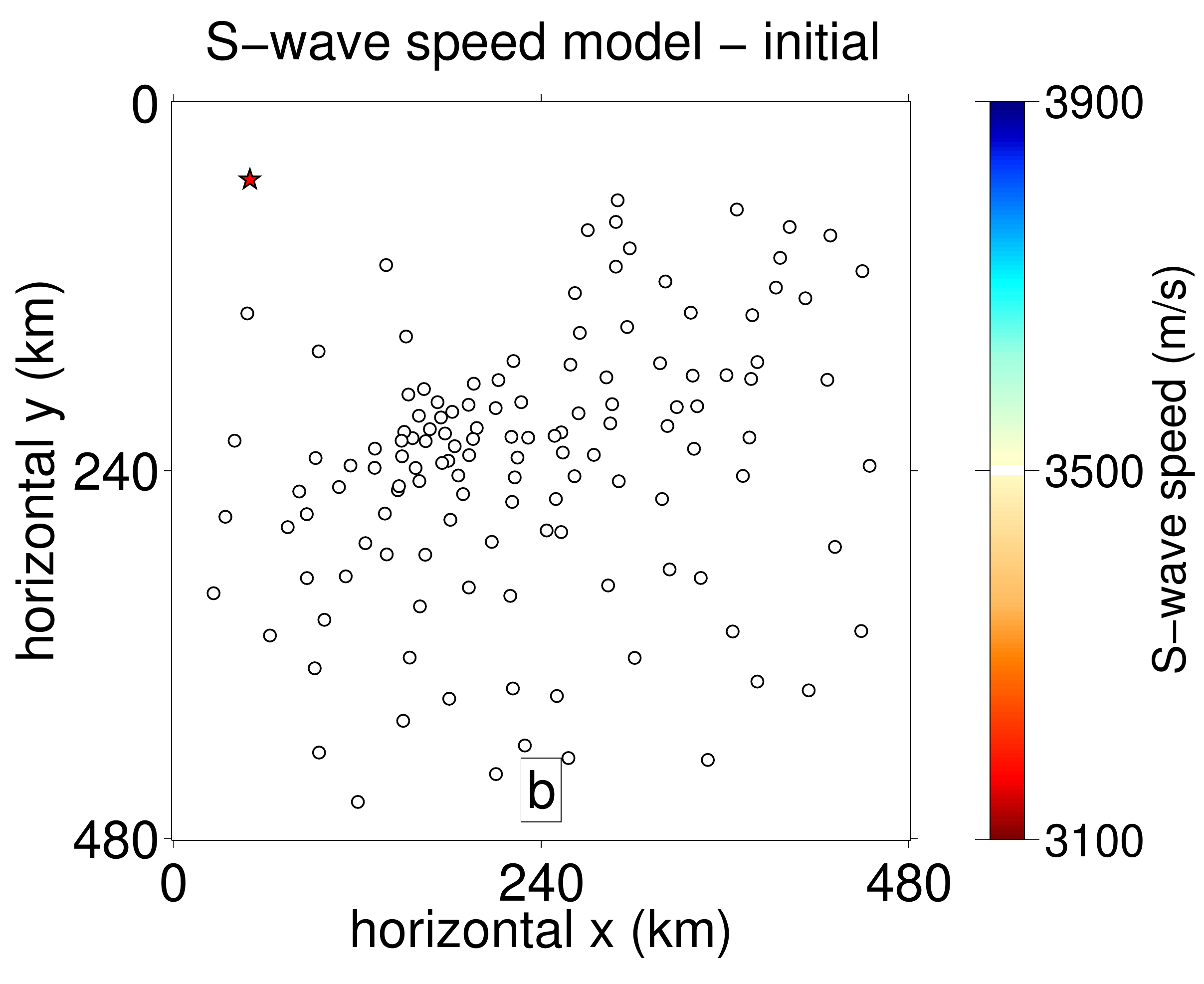}}
  \caption[Experiment~III: model configuration]{\label{fig:Tape_group3_model}
    Experiment~III: model configuration. One source (star) is placed
    near the top left corner and 132 stations (circles) are scattered
    about, off to the lower-right corner of the model
    space. (a)~Target wavespeed model. (b)~Initial model.}
\end{figure*}

\begin{figure*}\centering
  \rotatebox{-90}{\includegraphics[width=\model]{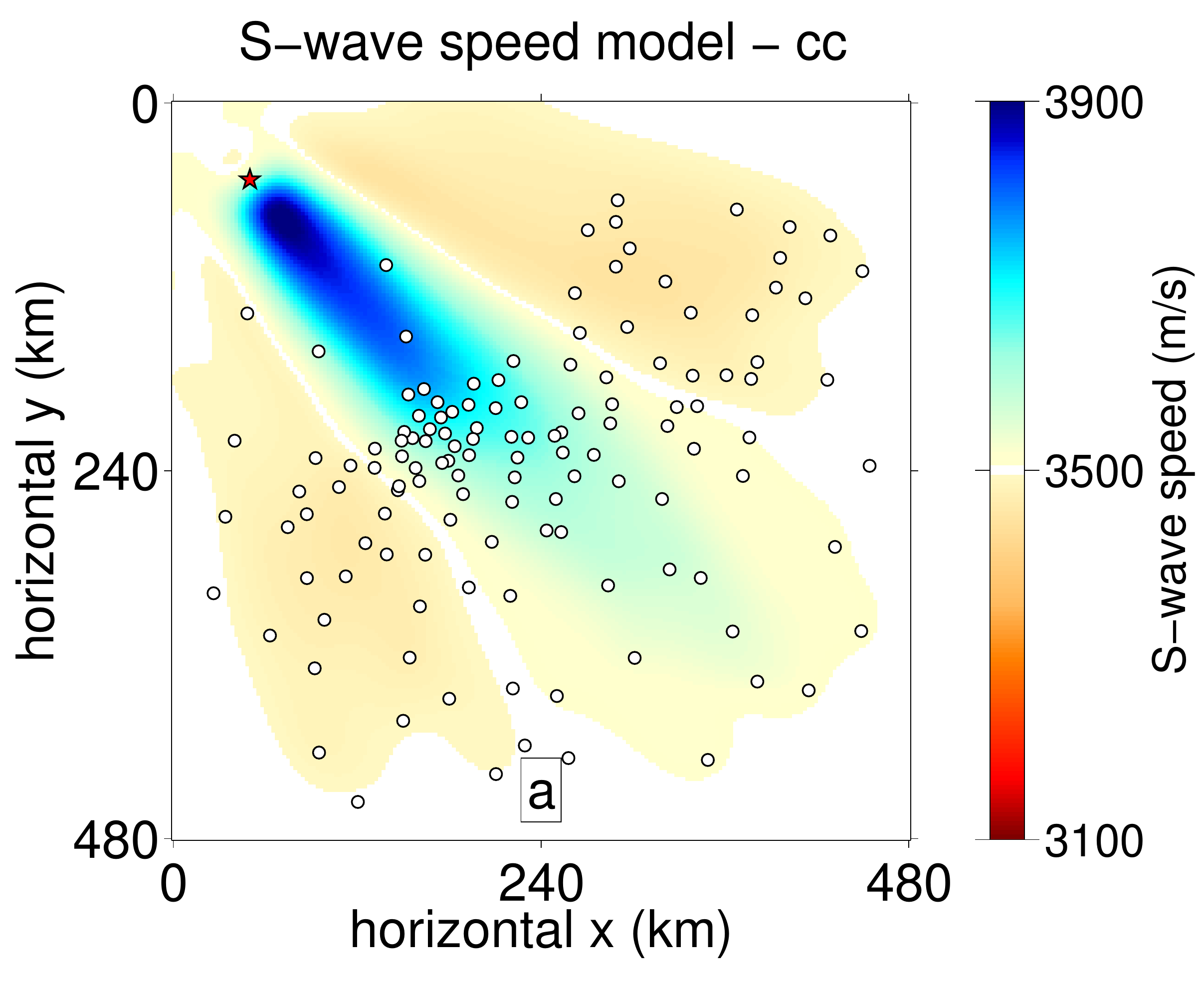}}\hspace{2em}
  \rotatebox{-90}{\includegraphics[width=\model]{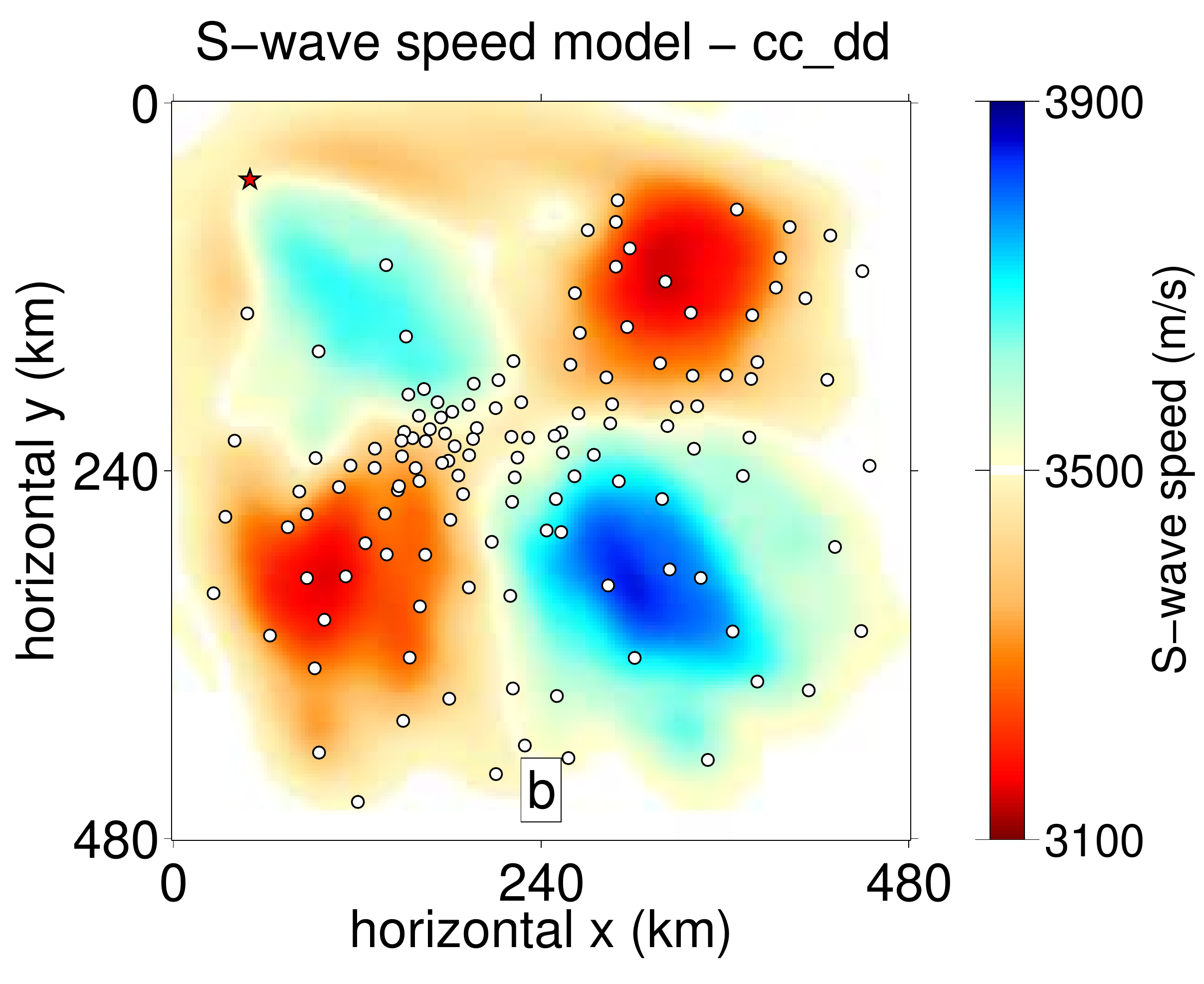}}
  \caption[Experiment~III: inversion
    results]{\label{fig:Tape_test7_model} Experiment~III: inversion
    results. Estimated shear-wave speed models with exact source-time
    function but different origin time. (a)~Using conventional
    cross-correlation traveltime measurements
    (10~iterations). (b)~Using double-difference
    cross-correlation traveltime measurements (7~iterations).}
\end{figure*}

In Experiment~III, we consider one earthquake and an array of stations
inspired by \cite{Tape+2007}, using location information from 132
broadband receivers in the Southern California Seismic Network
(SCSN). The target and initial models and the station-receiver geometry
are shown in Fig.~\ref{fig:Tape_group3_model}. An iterative inversion
based on conventional cross-correlation misfit measurements and
eq.~(\ref{kp}) results in the estimated structure shown in in
Fig.~\ref{fig:Tape_test7_model}a. An inversion based on
double-difference cross-correlation misfit measurements and
eq.~(\ref{kp2}) is shown in Fig.~\ref{fig:Tape_test7_model}b. The
final models show great differences between them. The misfit gradient
obtained with the conventional approach provides very limited
structural information, and the `final' model displays prominent
`streaks' between the stations and the source. In contrast, the
double-difference method is largely free from such artifacts and
yields significant detail on structural variations in the area under
consideration. \enlargethispage{2em}

Our checkerboard synthetic experiments were designed to demonstrate
the ease and power of the double-difference concept in adjoint
tomography, from making the measurements to calculating the kernels,
to inverting for the best-fitting models. It is immediately apparent
that the double-difference approach provides powerful interstation
constraints on seismic structure, and is less sensitive to error or
uncertainty in the source. Even using just a single earthquake, the
double-difference method enables the iterative spectral-element
adjoint-based inversion procedure to capture essential details of the
structure to be imaged.

In experiments (not shown here) where the starting model is
homogeneous but does not have the same mean wavespeed as the target
model, the double-difference approach manages to recover much of the
absolute and relative structure interior to the array, although
regions outside of the region encompassing the bulk of the stations
remain unupdated. In such cases, a hybrid method that combines both
traditional and double-difference measurements brings improvements
throughout the imaged region. 


\subsection{Realistic phase speed input models, global networks}

\begin{figure*}
  \includegraphics[angle=-90,width=\models,trim=0cm 0cm 1.5cm 0cm, clip]
  {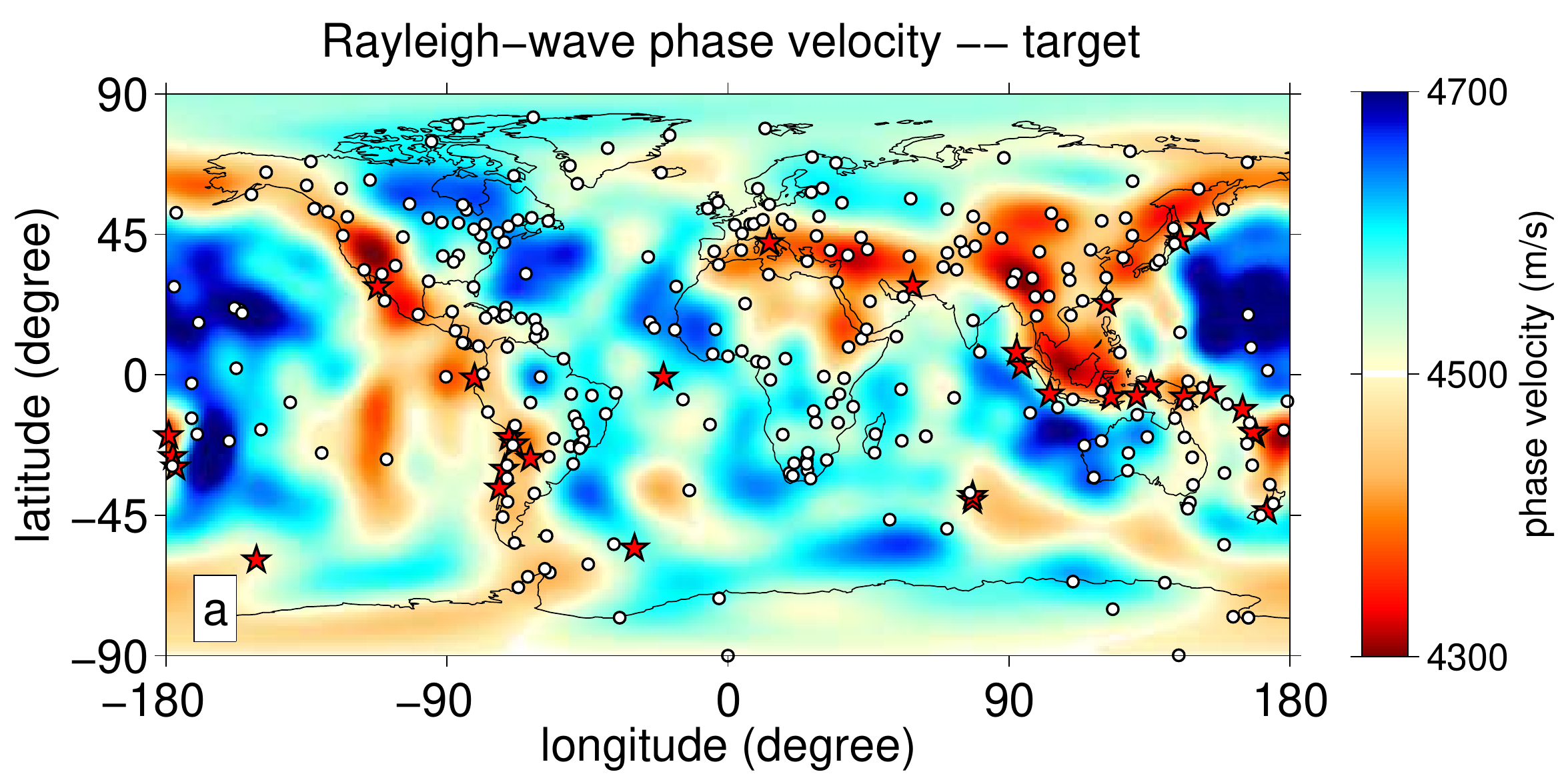}
  \includegraphics[angle=-90,width=\models,trim=0cm 0cm 1.5cm 0cm, clip]
  {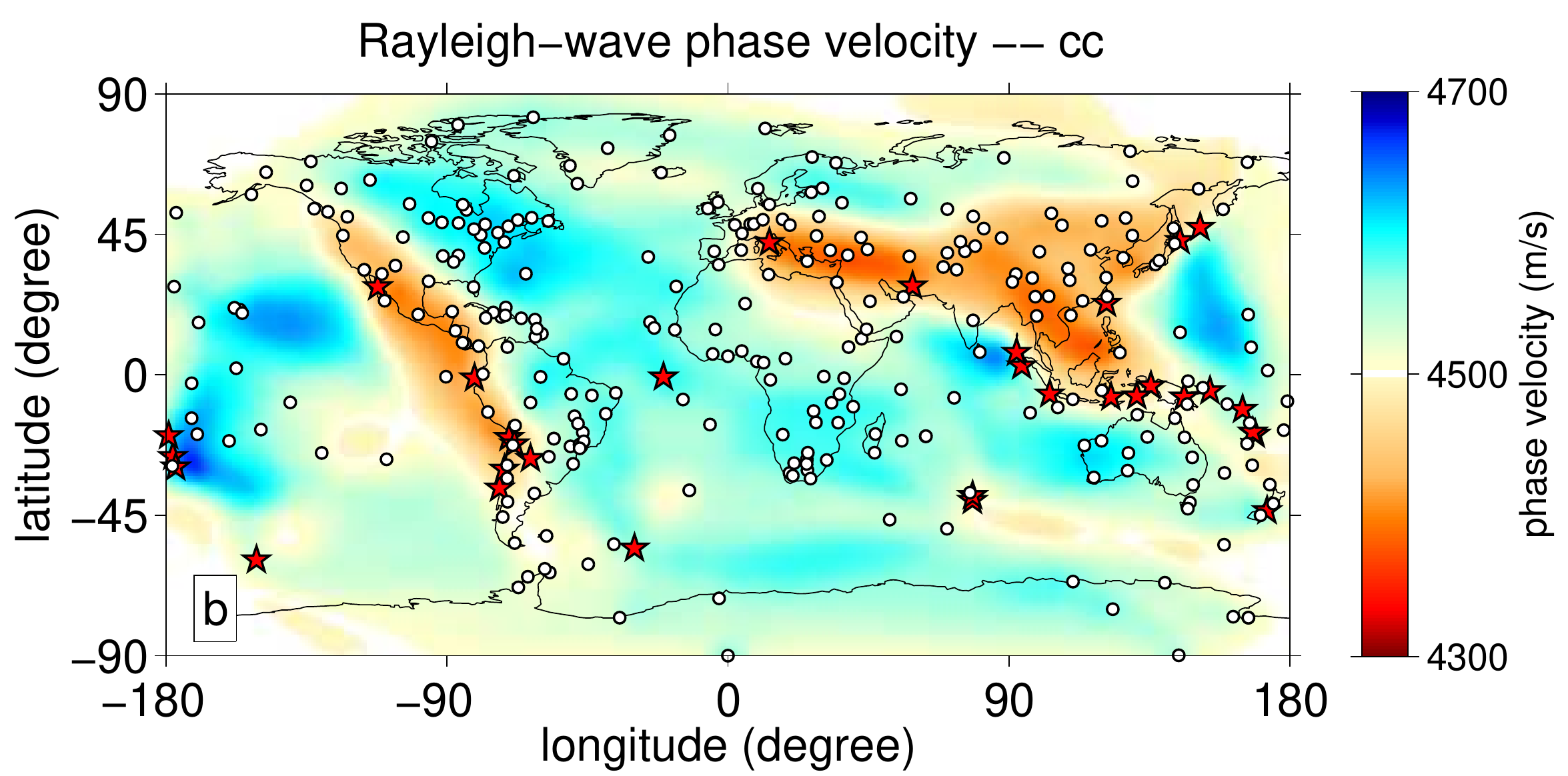}
 \includegraphics[angle=-90,width=\models]{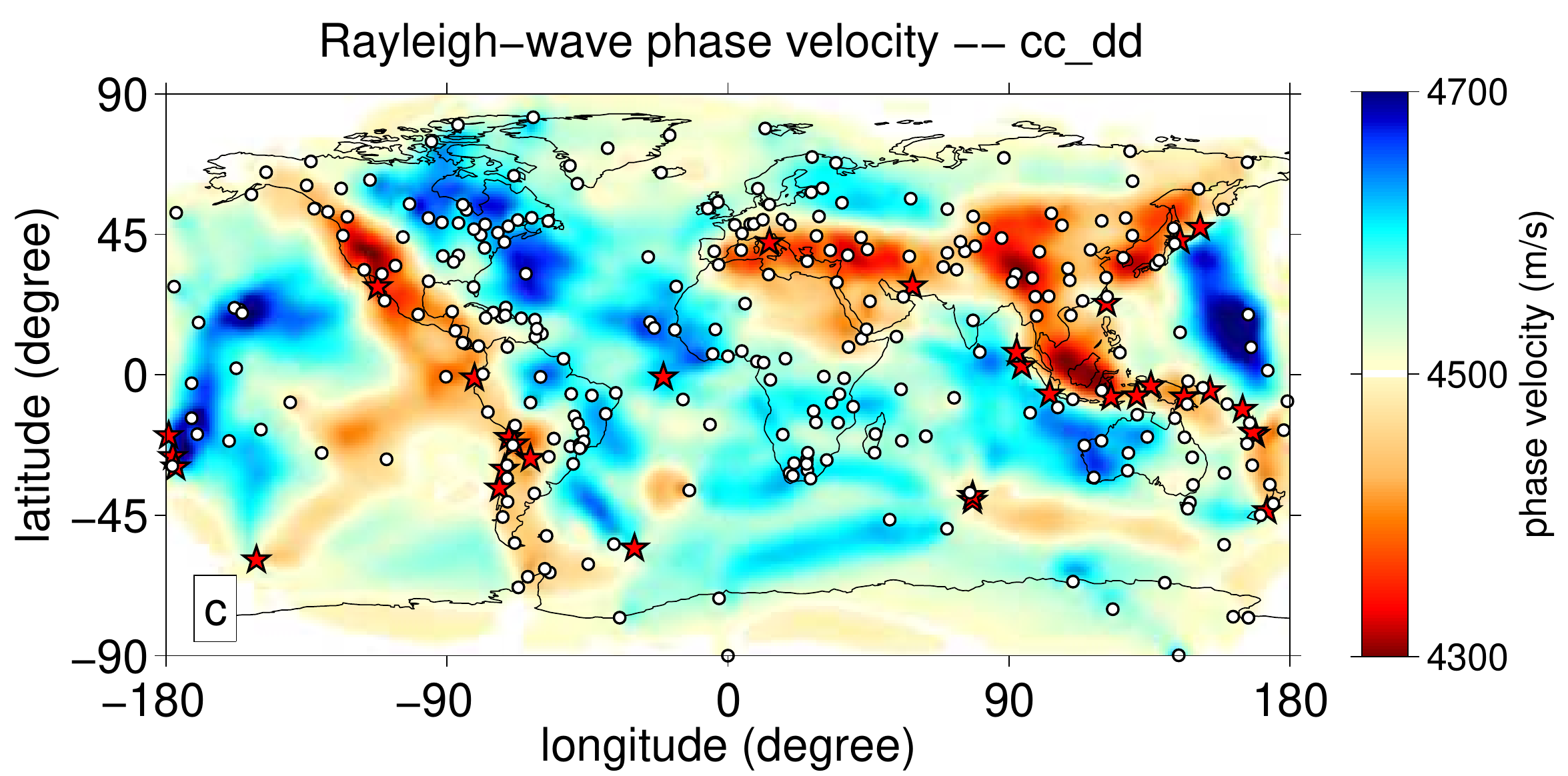}
 \caption[Experiment~IV: model and inversion
   results]{\label{global_test1} Experiment~IV: model and inversion
   results. (a)~The two-dimensional target model, inspired by
   \cite{Trampert+95}. The stars are 32 earthquakes. The open
   circles represent 293 global network stations. The initial model is
   homogeneous. (b)~Result after inversion with the conventional
   cross-correlation traveltime adjoint method
   (16~iterations). (d)~Inversion result from the double-difference
   cross-correlation traveltime adjoint method (13~iterations).}
\end{figure*}

To test realistic length scales in a global tomography model, we use
the global phase velocity map of \cite{Trampert+95} at periods between
40~s and 150~s, which we mapped to a simplified, unrealistic geometry
of a plane with absorbing boundaries. In Experiment~IV we model
membrane surface waves in this model, as shown in
Fig.~\ref{global_test1}a. The 32 selected earthquakes are shown by red
stars, and white circles depict 293 station locations from the global
networks.

\enlargethispage{1em}

We use the \texttt{SPECFEM2D} code to model the wavefield. We use 40
elements in latitude and 80 elements in longitude with an average of
4.5$^\circ$ in element size in each direction.  The rescaling
corresponds to dimensions of one meter per latitudinal degree
($^\circ$) of the original. The source wavelet is a Ricker function
with 260~Hz dominant frequency, which should be able to resolve
structure down to a scale length of approximately 6.6$^\circ$. The
sampling rate is 26~$\mu$s and the recording length 125~ms.


The initial model is homogeneous with a phase speed of 4500~m/s.
Using single-station cross-correlation traveltime measurements, the
inversion yields a model that has very limited resolution, shown in
Fig.~\ref{global_test1}b. From double-difference cross-correlation
measurements between stations, the inversion reveals structural
variations with much greater detail, as shown in
Fig.~\ref{global_test1}c. The conventional tomographic method appears
to over-emphasize source-rich areas, which is a result of the
summation of all sensitivities along the ray paths emitted from the
common source, while obscuring areas with a low density of overlapping
ray paths. On the other hand, the double-difference approach, by
joining all possible pairs of measurements, does cancel out much of
the shared sensitivity and, instead, illuminates wavespeed structure
locally among station clusters. With a more realistic geometry on the
sphere, we would expect to recover structure in the polar regions and
at longitude $180^\circ$, as well as gain slight improvements for the
interior structure.

\subsection{Realistic phase speed input models, regional seismic arrays}

Perhaps the greatest contribution of the double-difference tomographic
technique as we developed it here will lie in its application to
regional seismic arrays. The pairwise combination of seismic
measurements should enable imaging local structures with high spatial
resolution.

\begin{figure*}
  \includegraphics[angle=-90,width=\arrayss,trim=0cm 0cm 0cm 3.6cm,clip]
  {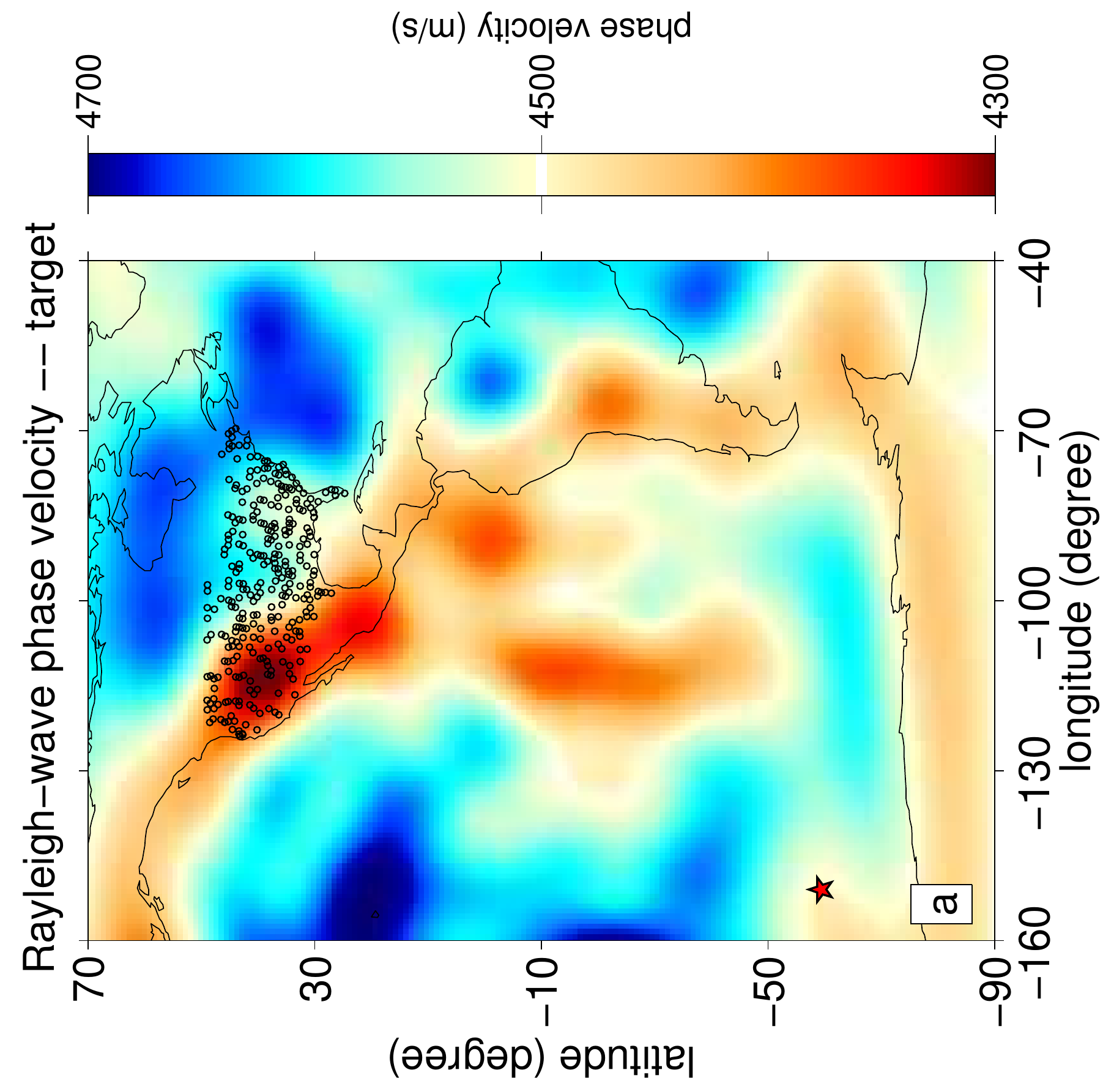}
  \includegraphics[angle=-90,width=\arrayss,trim=0cm 0cm 0cm 3.6cm,clip]
  {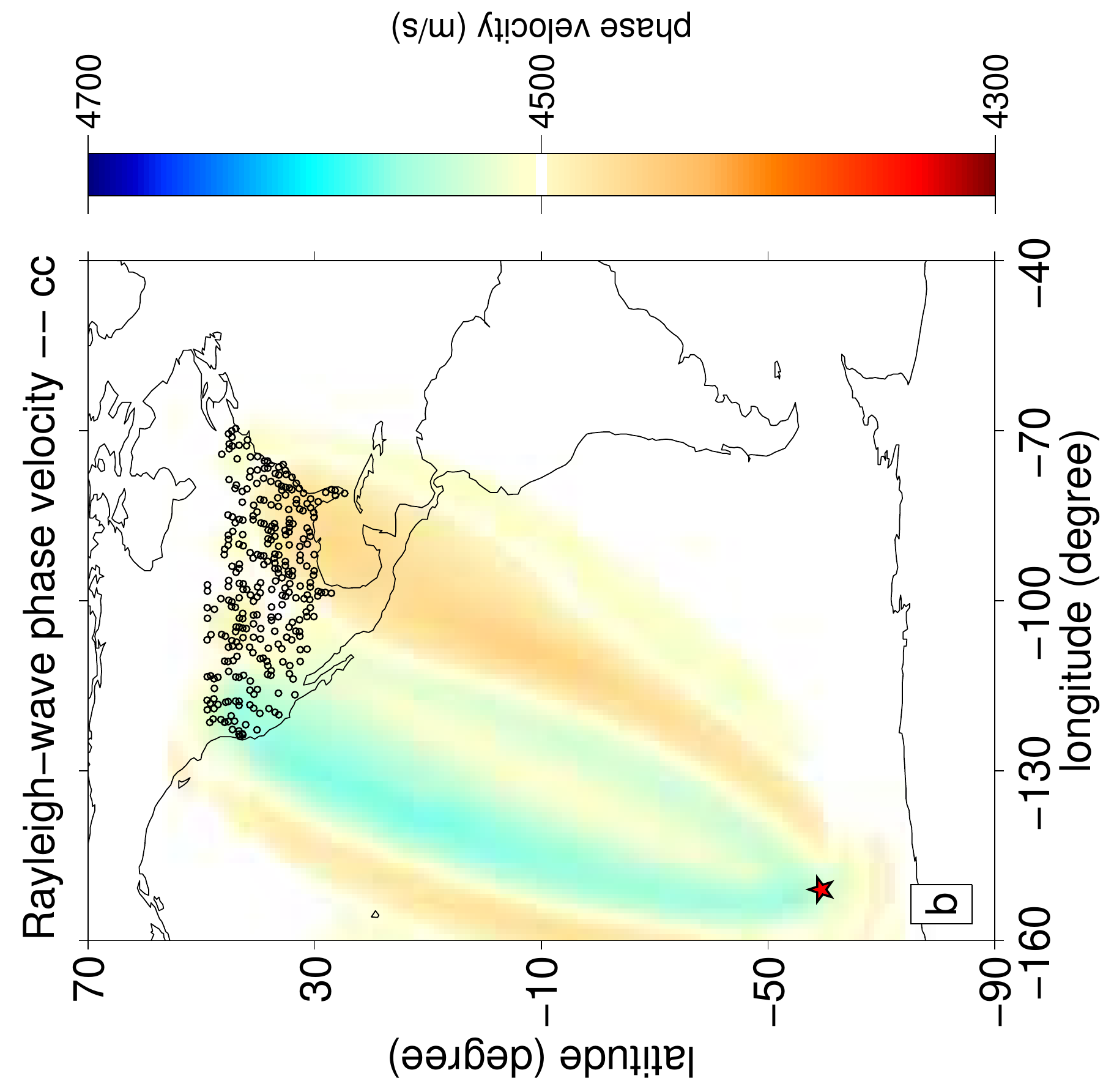}
  \includegraphics[angle=-90,width=\arrays]
  {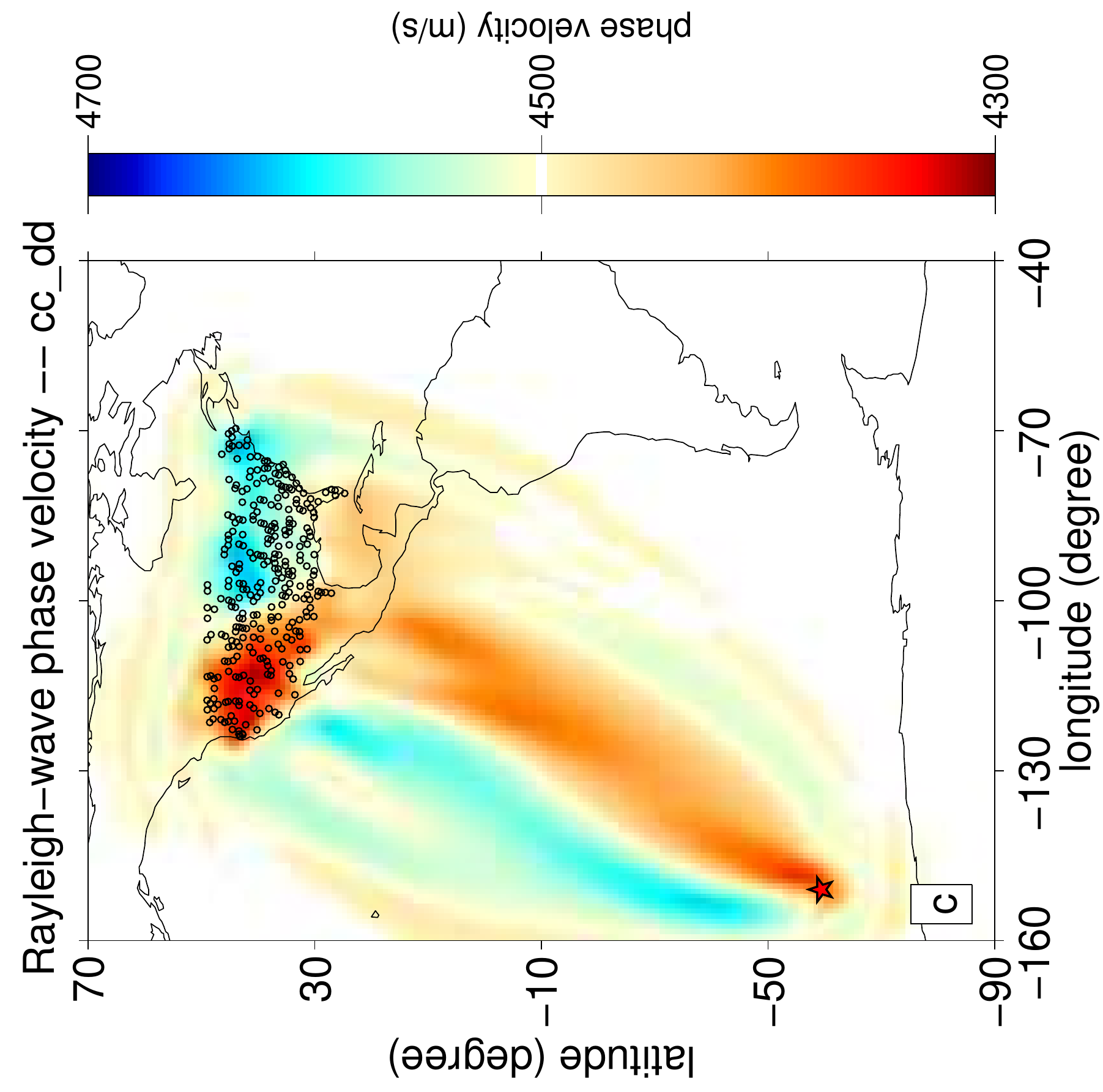}
   \caption[Experiment~V: model and inversion
     results]{\label{global_test2} Experiment~V: model and inversion
     results. (a)~The two-dimensional target model, a portion of the
     model shown in Fig.~\ref{global_test1}a. The star is the single
     earthquake source used in this study. The open circles represent
     300 stations from the USArray network. (b)~Inversion result with
     the conventional cross-correlation traveltime adjoint method
     (6~iterations). (d)~Inversion result using the double-difference
     method (6~iterations).}
\end{figure*}

In Experiment~V we continue the test of the synthetic model used in
the global-scale Experiment~IV, but this time with station located at
the sites of the Transportable Array (TA) system in the United States.
A view of the target model with the source-station geometry marked is
shown in Fig.~\ref{global_test2}a. The average station spacing is
about 0.7$^\circ$.  The double-difference technique is expected to
resolve scale lengths comparable to the station spacing, which is much
smaller than the scale of structural variations in this model. We
randomly selected 300 stations from the USArray station pool for
tomographic testing using membrane surface waves as in all prior
experiments.

The initial model is homogeneous. The inversion result in
Fig.~\ref{global_test2}b was obtained from the adjoint tomography with
conventional cross-correlation traveltime measurements recorded at all
stations. The model resolves little more than the average wavespeed
structure along the common ray path between the earthquake and the
array. In contrast, the double-difference method reveals substantial
details of the local structure beneath the array, and the
perturbations along the direct path are also reasonable. The intrinsic
sensitivity of differential measurements will make the
double-difference technique ideal for the high-resolution
investigation of well-instrumented areas with limited natural seismic
activity, or where destructive sources are not an option.

\section{C~O~N~C~L~U~S~I~O~N~S}

In conventional traveltime tomography, source uncertainties and
systematic errors introduce artifacts or bias into the estimated
wavespeed structure. In the double-difference approach, measurements
made on pairs of stations that share a similar source and similar
systematic effects are differenced to partially cancel out such
consequences. Both theoretically and in numerical experiments, we have
shown that the `double-difference' approach produces reliable misfit
kernels for adjoint tomography, despite potentially incorrect source
wavelets, scale factors, or timings used in synthetic modeling.

Furthermore, double-difference adjoint tomography is an efficient
technique to resolve local structure with high resolution. By pairing
seismograms from distinct stations with a common source, the technique
yields sharp images in the area local to the stations. Those areas are
often plagued by smearing artifacts in conventional tomography,
especially when limited numbers of earthquakes are available nearby.
This opens a wide range of promising applications in seismic array
analysis, and especially in areas where the use of active sources is
restricted. Fine-scale structure within the array area can be resolved
even from a single earthquake.

The double-difference algorithm can be simply implemented into a
conventional adjoint tomography scheme, by pairing up all regular
measurements. The adjoint sources necessary for the numerical gradient
calculations are obtained by summing up contributions from all pairs,
which are simultaneously back-propagated to calculate the differential
misfit kernels using just one adjoint simulation per iteration.

\section{A~c~k~n~o~w~l~e~d~g~e~m~e~n~t~s}

YOY expresses her thanks to Min Chen, Paula Koelemeijer, Guust Nolet,
and Youyi Ruan for discussions. This research was partially supported
by NSF grants EAR-1150145 to FJS and EAR-1112906 to JT, and by a
Charlotte Elizabeth Procter Fellowship from Princeton University to
YOY. We thank Carl Tape and Vala Hj\"orleifsd\'ottir for sharing
unpublished notes related to their published work which we cite. The
Associate Editor, Jean Virieux, Carl Tape, and one other, anonymous,
reviewer are thanked for their thoughtful and constructive comments,
which improved our manuscript. Our computer codes are available from 
\url{https://github.com/yanhuay/seisDD}.


\section{A~P~P~E~N~D~I~X:{\hsps}D~O~U~B~L~E~-~D~I~F~F~E~R~E~N~C~E{\hsps}M~U~L~T~I~T~A~P~E~R{\hsps}M~E~A~S~U~R~E~M~E~N~T~S}

In the main text we focused on measuring traveltime shifts via
cross-correlation, a time-domain method that does not offer frequency
resolution beyond what can be achieved using explicit filtering
operations prior to analysis. Such an approach is implicitly
justified for the inversion of body waves, but for strongly
dispersive waves such as surface waves, we recommend making
frequency-dependent measurements of phase and amplitude using the
multitaper method \cite[]{Park+87a,Laske+96,Zhou+2004,Tape+2009}. In
this section we discuss its application to the double-difference
framework. Our derivations borrow heavily from
\cite{Hjorleifsdottir2007}. They are purposely `conceptual' for
transparency, and to remain sufficiently flexible for a variety of
practical implementation strategies.

\subsection{Multitaper measurements of differential phase and
  amplitude}

For notational convenience we distinguish time-domain functions such
as $s(t)$ or $d(t)$ from their frequency-domain counterparts $s\ofo$
and $d\ofo$ only through their arguments $t$ for time, and $\omega$
for angular frequency. For two stations $i$ and $j$ (always
subscripted and thus never to be confused with the imaginary number
$i=\sqrt{-1}$), the Fourier-transformed waveforms are assumed to be
of the form \cite[e.g.,][]{Dziewonski+72,Kennett2002b}
\begin{subequations}
  \label{bli}\begin{align}
    \label{eq:FT_si}
    s_i\ofo &= A_i\syn\ofo \,e^{-i\omega \,t_i\syn\ofo } 
    \also s_j\ofo  = A_j\syn\ofo \,e^{-i\omega\, t_j\syn\ofo }
    ,\\
    \label{eq:FT_di}
    d_i\ofo &= A_i\obs\ofo \,e^{-i\omega \,t_i\obs\ofo}
    \also d_j\ofo  = A_j\obs\ofo \,e^{-i\omega\, t_j\obs\ofo }
    .
  \end{align}
\end{subequations}
We define the differential frequency-dependent `phase' (compare with
eqs~\ref{eq:cc}) and `logarithmic amplitude' anomalies as
\begin{subequations}
  \label{pa}
  \begin{align}
    \label{eq:def_tau_syn_ij}
    \dtss\ofo &= t_i\syn\ofo-t_j\syn\ofo
    \also
    \DlnA\sij\syn\ofo = \ln\left[\frac{A_i\syn\ofo}{A_j\syn\ofo}\right]
    ,
    \\ 
    \label{eq:def_tau_obs_ij}
    \dtso\ofo &= t_i\obs\ofo
    -t_j\obs\ofo 
    \also
    \DlnA\sij\obs\ofo = \ln\left[\frac{A_i\obs\ofo}{A_j\obs\ofo}\right]
    .
  \end{align}
\end{subequations}
We adopt the idealized viewpoint that a trace recorded at a
station~$i$ may be mapped onto one made at a station~$j$ \cite[perhaps
  after specific phase-windowing, see][and/or after time-shifting and scaling by a
  prior all-frequency cross-correlation measurement]{Maggi+2009} via a
`frequency-response' or `transfer' function, $\cT\sij\ofo$,
\begin{subequations}\label{blu}
  \begin{align}\label{eq:tf_syn} 
    s_i\ofo &=\cTfs\ofo \,s_j\ofo,\\
    \label{eq:tf_obs} 
    d_i\ofo &=\cT\sij\obs\ofo \,d_j\ofo
    .
  \end{align}
\end{subequations}
Combining eqs~(\ref{bli})--(\ref{blu}), we can rewrite the theoretical
`transfer functions' of our model as
\begin{subequations}
  \label{junk}
  \begin{align}
    \label{eq:tf_syn_ij}
    \cTfs\ofo&=
    e^{\DlnA\sij\syn\ofo -i\omega \dtss \ofo }
    ,\\
    \label{eq:tf_obs_ij} 
    \cT\sij\obs\ofo&=
    e^{\DlnA\sij\obs\ofo -i\omega \dtso \ofo}
    .
  \end{align}
\end{subequations}
We have hereby defined $\dtss\ofo $ and $\dtso\ofo $ as the
frequency-dependent phase differences between $s_i(t)$ and $s_j(t)$,
and $d_i(t)$ and $d_j(t)$ respectively, and $\DlnA\sij\syn\ofo $
and $\DlnA\sij\obs\ofo$ with their frequency-dependent amplitude
differences.

The linear time-invariant model~(\ref{blu}) is far from perfect
\cite[for a time-and-frequency dependent viewpoint, see,
  e.g.,][]{Kulesh+2005,Holschneider+2005}: the transfer functions
must be `estimated'. Following \cite{Laske+96}, we begin by
multiplying each of the individual time-domain records $s_i(t)$ and
$d_i(t)$ by one of a small set of $K$ orthogonal data windows denoted
$h^{T\Omega}_k(t)$ \cite[here, the prolate spheroidal functions
  of][]{Slepian78}, designed for a particular record length~$T$ and
for a desired resolution angular-frequency half-bandwidth~$\Omega$
\cite[]{Mullis+91,Simons+2015}, and Fourier-transforming the result
to yield the sets
\begin{equation}
  s_i^k\ofo, s_j^k\ofo \also
  d_i^k\ofo, d_j^k\ofo \where k=1,\ldots,\left\lfloor \frac{T\Omega}{\pi}
  \right\rfloor-1
  .
\end{equation}
At frequencies separated by about $2\Omega$, any differently
`tapered' pairs $s^k,s^{l}$ or $d^k,d^{l}$ will be almost perfectly
uncorrelated statistically \cite[]{Percival+93}. Using all of them,
the transfer function in eq.~(\ref{eq:tf_syn_ij}) can be estimated in
the least-squares sense \cite[][Ch.~9]{Bendat+2000} as minimizing
the linear-mapping `noise' power associated with
eq.~(\ref{eq:tf_syn}),
\begin{equation}\label{test}
  \Tfs\ofo =
  \frac{\sumk s_i^k\ofo \big[s^k_j\ofo \big]^*}
       {\sumk s^k_j\ofo\big[s^k_j\ofo \big]^*}=
       \argmin_{\cTfs}\sumk \left\{s_i^k\ofo-\cTfs\ofo  s^k_j\ofo \right\}^2 
       .
\end{equation}
The superscripted asterisk ($^*$) denotes complex conjugation. We note
here for completeness that the associated `coherence' estimate is
\begin{equation}\label{coh}
  \gamma^2\sij\ofo=
  \frac{\left|\sumk s_i^k\ofo \big[s^k_j\ofo \big]^*\right|^2}
       {\sumk s^k_i\ofo\big[s^k_i\ofo \big]^*
         \suml s^l_j\ofo\big[s^l_j\ofo \big]^*}
       .
\end{equation}
Switching to polar coordinates we write the estimate of the transfer
function in terms of its  `gain factor' and `phase angle'
contributions,
\begin{equation}\label{polar}
  \Tfs\ofo=\big|\Tfs\ofo\big|e^{-i\phi\sij\syn(\omega)}
  ,
\end{equation}
where
\begin{equation}\label{jobo}
  \big|\Tfs\ofo\big|=
  \left(\big\{\Im\big[\Tfs\ofo\big]\big\}^2+
  \big\{\Re\big[\Tfs\ofo\big]\big\}^2\right)^{1/2}
  \also
  \phi\sij\syn(\omega)=-\atan
  \left\{
  \frac
      {\Im\big[\Tfs\ofo\big]}
      {\Re\big[\Tfs\ofo\big]}
      \right\}
      .
\end{equation}
Uncertainties on the phase angle, gain factor, and coherence estimates
are approximately \cite[]{Bendat+2000,Simons+2003a}
\begin{equation}\label{thunc}
  \var\{\phi\sij\syn(\omega)\}
  \approx\frac{1-\gamma^2\sij\ofo}{2K\hspso\gamma^2\sij\ofo}
  ,\qquad
  \var\big\{|\Tfs|\big\}
  \approx\frac{1-\gamma^2\sij\ofo}{2K\hspso\gamma^2\sij\ofo}\big|\Tfs\big|^2
  \also
  \var\big\{\gamma^2\sij\ofo\}
  \approx
  \frac{\big[1-\gamma^2\sij\ofo\big]^2}{K/2}\gamma^2\sij\ofo
  ,
\end{equation}
where we have silently subscribed to the established practice of
confusing the estimated (on the left hand side of the expressions)
from the `true' quantities (on the right). Alternative estimates of
the quantities in eqs~(\ref{test}) and~(\ref{coh}) could be obtained
through `jackknifing' \cite[]{Chave+87,Thomson+91}. We follow
\cite{Laske+96} in using eq.~(\ref{test}) to estimate the transfer
function, but we do estimate its variance through the jackknife, which
will be conservative \cite[]{Efron+81}. More detailed distributional
considerations are provided by \cite{Carter87}. The utility of
eq.~(\ref{thunc}) lies in the interpretation that highly `coherent'
(in the sense of eq.~\ref{coh}) waveforms estimated from a relatively
large number of multitapered Fourier-domain `projections' will be,
quite generally, very precise.

Through eqs.~(\ref{eq:tf_syn_ij}) and~(\ref{polar}), finally, we
arrive at our estimates for the frequency-dependent differential phase
and amplitude measurements on the synthetic time-series $s_i(t)$ and
$s_j(t)$, namely
\begin{subequations}
  \label{stuff}
  \begin{align}
    \label{eq:tau_syn_ij}
    \Delta t\syn\sij\ofo &=
    \phi\sij\syn(\omega)/\omega
    \also
    \DlnA\sij\syn\ofo =
    \ln\big|{\Tfs\ofo}\big|
    . 
  \end{align}
  Exchanging the synthetic time-series for the observed ones in the
  above expressions, $s\leftrightarrow d$, accomplishes the desired
  switch in the superscripted $\mathrm{syn}\leftrightarrow\mathrm{obs}$,
  whence
  \begin{align}
    \label{eq:tau_obs_ij}
    \Delta t\obs\sij\ofo &=
    \phi\sij\obs(\omega)/\omega
    \also
    \DlnA\sij\obs\ofo =
    \ln\big|{T\sij\obs\ofo}\big|
    . 
  \end{align}
\end{subequations}
Error propagation will turn eqs~(\ref{thunc}) into the relevant
uncertainties on the differential phase and amplitude measurements in
eqs~(\ref{stuff}). 


The logarithmic amplitude measurements in eqs~(\ref{stuff}) are not
skew-symmetric under a change of the station indices, since
eq.~(\ref{test}), the estimated transfer function from station~$i$ to
$j$, is not simply the inverse of the transfer function from
station~$j$ to $i$, as indeed it need not be: the modeling
imperfection in mapping the seismograms is not simply
invertible. Instead we find from eqs~(\ref{test})--(\ref{coh}) that
$T\sij T\sji=\gamma^2\sij$, suggesting that the required symmetry
prevails only at frequencies at which the waveforms are highly
coherent. A transfer-function estimate alternative to eq.~(\ref{test})
might thus be $T\sij/|\gamma\sij|$, as under this scenario a switch in
the indices would indeed lead to $\DlnA\sji$ being equal to
$-\DlnA\sij$ under all circumstances of mutual coherency. In the
least-squares interpretation we would then be minimizing the `noise'
power after mapping one seismogram onto another subject to weighting
by the absolute coherence to avoid mapping frequency components at
which the signals are incoherent. We do not make this choice \cite[or
  indeed, any other of many possible options listed by,
  e.g.,][]{Knapp+76} here, but we will revisit the issue of symmetry
later in this Appendix.

If the seismogram pairs had been pre-aligned and pre-scaled using
conventional cross-correlation, perhaps to improve the stability of
the frequency-dependent phase and amplitude expressions~(\ref{jobo}),
these time-domain, frequency-independent, measurements would need to
be added to the traveltime and amplitude expressions~(\ref{stuff}). If
specific time windows of interest were being involved, all of the
subsequent formalism would continue to hold but, naturally, the
effective number of measurement pairs would increase
proportionally. And of course, the double-difference multitaper method
could follow any number of prior inversions designed to bring the
starting model closer to the target; in this context we mention our
own work on wavelet multiscale waveform-difference,
envelope-difference and envelope-cross-correlation adjoint techniques
\cite[]{Yuan+2014a,Yuan+2015}.

\subsection{Double-difference multitaper measurements, misfit functions, and
  their gradients}

Making differential measurements between stations resolves the
intrinsic ambiguity in the attribution of the phase and amplitude
behavior of individual seismic records to source-related,
structure-induced or instrumental factors \cite[]{Kennett2002b}, and
for this reason has been the classical approach for many decades
\cite[]{Dziewonski+69,Dziewonski+72}. But rather than using the
differential measurements for the interrogation of subsurface
structure on the interstation portion of the great circle connecting
one source to one pair of receivers, we now proceed to making, and
using, differential measurements between all pairs of stations into a
tool for adjoint tomography. We take our cue from
eq.~(\ref{eq:cc_DD}) and use eqs~(\ref{stuff}) to define the
`double-difference' frequency-dependent traveltime and amplitude
measurements made between stations $i$ and $j$, namely
\begin{subequations}
  \begin{align}
    \dd t\sij\ofo&=\dtss\ofo -\dtso\ofo
    \also
    \DDlnA\sij\ofo=\DlnA\sij\syn\ofo -\DlnA\sij\obs\ofo 
    .
  \end{align}
\end{subequations}

As in eq.~(\ref{eq:misfit_CC_DD}), we incorporate those into a phase
and an amplitude misfit function defined over the entire available
bandwidth and for all station pairs, i.e.
\begin{subequations}
  \label{ddmisfit}
  \begin{align}\label{eq:misfit_MTP_DD}
    \chi\PP\DD&= \half\sumh\sumo  W\PP\ofo
    \big[\dd t\sij\ofo \big]^2,\\
    \label{eq:misfit_MTQ_DD}
    \chi\A\DD&=\half\sumh\sumo  W\A\ofo  
    \big[\DDlnA\sij\ofo \big]^2
    ,
  \end{align}
\end{subequations}
with $W\PP\ofo$ and $W\A\ofo$ appropriate weight functions. The
weighted sum over all frequencies is meant to be respectful of the
effective bandwidth of the estimates derived via eq.~(\ref{test})
from seismic records $s_i(t)$, $s_j(t)$ and $d_i(t)$, $d_j(t)$
multitapered by the Slepian window set $h^{T\Omega}_k(t)$,
$k=1,\ldots,\left\lfloor T\Omega/\pi \right\rfloor-1$. The multitaper
procedure implies `tiling' the frequency axis over the entire
bandwidth in pieces of phase and amplitude `information' extracted
over time portions of length~$T$, centered on individual frequencies
$\omega$, that are approximately uncorrelated when spaced $2\Omega$
apart \cite[]{Walden90a,Walden+95}. The simplest form of the weight
function is thus $W\PP\ofo=W\A\ofo=2\Omega$, for a discrete set of
angular frequencies $\omega=\Omega,3\Omega,\ldots$ until the complete
signal bandwidth is exhausted, tapering slightly at the edges of the
frequency axis to smoothly contain the full bandwidth of the
seismograms. Alternatively, to serve the purpose of filtering out
frequencies containing incoherent energy
\cite[][Ch.~6]{Bendat+2000}, $W\PP\ofo$ and $W\A\ofo$ can be the
inverse of the phase and amplitude variances~(\ref{thunc}), and, if
subsampling is undesirable, they should reflect the inverse full
covariances of the estimates between frequencies
\cite[]{Percival+93}.

As in eq.~(\ref{eq:derivative_misfit_CC_DD}), the derivatives of the
differential objective functions in eqs~(\ref{ddmisfit}) are
\begin{subequations}\label{bliblo}
  \begin{align}
    \delta \chi\PP\DD
    &=\label{eq:derivative_MTP_DD}
    \sumh \sumo  W\PP\ofo   
          [\dd t\sij\ofo]
          \,\delta\Delta t\syn\sij\ofo
          ,\\
          \delta \chi\A\DD 
          &=\label{eq:derivative_MTQ_DD}
          \sumh \sumo  W\A\ofo   
                [\DDlnA\sij\ofo]
                \,\dDlnA\syn\sij\ofo
                ,
  \end{align}
\end{subequations}
where $\delta \Delta t\syn\sij\ofo$ is the perturbation of the
frequency-dependent differential traveltime~$\dtss\ofo$ due to a
model perturbation, and $\delta \DlnA\syn\sij\ofo$ the
perturbation of the frequency-dependent differential
amplitude~$\DlnA\sij\syn\ofo$. It is for the terms $\delta \Delta
t\syn\sij\ofo$ and $\delta \DlnA\syn\sij\ofo$ that we will
construct expressions leading to a new set of adjoint sources.

\subsection{Double-difference multitaper adjoint sources}

By taking perturbations to the formula in~(\ref{eq:tf_syn_ij}) in its
estimated (italicized) form, we have
\begin{align}\label{eq:tf_syn_perturb}
  \delta \Tfs\ofo  
  &=   \Tfs\ofo  \big[ \delta \DlnA\syn\sij\ofo  - i\omega \delta \Delta t\syn\sij\ofo  \big].
\end{align}
Dropping the argument that captures the dependence on frequency for
notational convenience, we use the product and chain rules for
differentiation to write the total derivative of eq.~(\ref{test}) as
\begin{align}\label{eq:tf_syn_mt_perturb}
  \delta \Tfs\ofs  
  &=\frac{1}{S\sjj}\sumk\, [s_j^k\ofs]^* \,\delta s_i^k\ofs  
  -\Tfs\ofs [s_j^k\ofs]^* \,\delta s_j^k\ofs
  +\left( s_i^k\ofs -\Tfs\ofs s_j^k\ofs \right)
  \,\delta [s_j^k\ofs]^*
  .
\end{align}
We introduced a symbol for the multitapered (cross-) power-spectral
density estimates
\begin{equation}
  S\sij=\sumk s^k_i\ofs [s^k_j\ofs]^*
  \with
  \Tfs=\fracd{S\sij}{S\sjj}
  \also
  S\sij^*=S\sji
  ,
\end{equation}
which of course upholds the relation eq.~(\ref{test}). Combining
eqs.~(\ref{eq:tf_syn_perturb})--(\ref{eq:tf_syn_mt_perturb}) and using
eq~(\ref{test}) again, we then obtain the perturbations
\begin{align}\label{eq:dt_perturb}
  \delta \Delta t\syn\sij\ofo
  &=\frac{i}{2\omega}
  \sumk\frac{[s_j^k\ofs]^*}{S\sij} \,\delta s_i^k\ofs 
  -\frac{s_j^k\ofs}{S\sij^*} \,\delta [s_i^k\ofs]^* 
  -\frac{[s_i^k\ofs]^*}{S\sij^*} \,\delta s_j^k\ofs  
  +\frac{s_i^k\ofs}{S\sij}\,\delta [s_j^k\ofs]^*
\end{align}
through straightforward manipulation, and, likewise,
\begin{align}\label{eq:dA_perturb}
  \dDlnA\syn\sij\ofo
  &=\half\sumk\frac{[s_j^k\ofs]^*}{S\sij} \,\delta s_i^k\ofs 
  +\frac{s_j^k\ofs}{S\sij^*} \,\delta [s_i^k\ofs]^*
  +\left(\frac{[s_i^k\ofs]^*}{S\sij^*}-2\frac{[s_j^k\ofs]^*}{S\sjj}\right) \delta s_j^k\ofs
  +\left(\frac{s_i^k\ofs}{S\sij}-2\frac{s_j^k\ofs}{S\sjj}\right) \delta [s_j^k\ofs]^*
  .
\end{align}
We shall associate the frequency-dependent partials that appear in
eqs~(\ref{eq:dt_perturb})--(\ref{eq:dA_perturb}) with the symbols
\begin{subequations}
  \begin{align}\label{ps}
    p_{i|j}^k\ofo&= \frac{i}{2\omega}\frac{[s_j^k\ofs]^*}{S\sij}
    \also        p_{j|i}^k\ofo= \frac{-i}{2\omega}\frac{[s_i^k\ofs]^*}{S\sij^*},
    \\\label{as}
    a_{i|j}^k\ofo&=\half\frac{[s_j^k\ofs]^*}{S\sij}
    \also        a_{j|i}^k\ofo=\half\frac{[s_i^k\ofs]^*}{S\sij^*}-\frac{[s_j^k\ofs]^*}{S\sjj},
  \end{align}
\end{subequations}
and their conjugates. Again, we remark on the lack of station symmetry
in the amplitude terms~(\ref{as}). However, it is noteworthy that, via
the Fourier identity $\pl_t\leftrightarrow i\omega$, the terms in the
expressions~(\ref{ps}) are identical (up to the sign) to the (first)
terms in eq.~(\ref{as}) if in the latter, the time-domain seismograms
are substituted for their time-derivatives. Consequently, we may
recognize eqs~(\ref{eq:dt_perturb})--(\ref{eq:dA_perturb}) as the
(differential) frequency-dependent (narrow-band) multitaper
generalization of the traveltime expressions~(\ref{dti})
and~(\ref{dtss}), familiar from \cite{Dahlen+2000}, and of their
amplitude-anomaly counterparts, see \cite{Dahlen+2002}. With these
conventions in place, the derivatives of the multitaper objective
functions in eq.~(\ref{bliblo}) are now given by
\begin{subequations}\label{blonki}
  \begin{align}\label{eq:blonki}
    \delta\chi\DD\PP=&
    \sumh\sumk \sumo W\PP\ofo [\dd t\sij\ofo]
    \left\{
    p_{i|j}^k\ofo \,\delta s_i^k\ofo
    +[p_{i|j}^k\ofo]^*  \,\delta [s_i^k\ofo]^*
    +p_{j|i}^k\ofo \,\delta s_j^k\ofo
    +[p_{j|i}^k\ofo]^* \,\delta [s_j^k\ofo]^*
    \right\}
    ,\\
    \delta\chi\DD\A=&
    \sumh\sumk \sumo W\A\ofo [\DDlnA\sij\ofo] 
    \left\{
    a_{i|j}^k\ofo \,\delta s_i^k\ofo
    +[a_{i|
        j}^k\ofo]^* \,\delta [s_i^k\ofo]^*
    +a_{j|i}^k\ofo \,\delta s_j^k\ofo
    +[a_{j|i}^k\ofo]^* \,\delta [s_j^k\ofo]^*
    \right\}
    .
  \end{align}
\end{subequations}
We introduce the time-domain functions, inverse-Fourier transformed from their
frequency-domain product counterparts, 
\begin{subequations}
  \begin{align}
    \mathscr{S}_{i|j}\oft&=
    \sumk h^{T\Omega}_k\oft\,\mathcal{F}^{-1}\!\left\{W\PP [\dd t\sij]
    \,p_{i|j}^k\right\}\!\oft,
    \also
    \mathscr{S}_{j|i}\oft=
    \sumk h^{T\Omega}_k\oft\,\mathcal{F}^{-1}\!\left\{W\PP [\dd t\sij]
    \,p_{j|i}^k\right\}\!\oft,
    \\
    \mathscr{A}_{i|j}\oft&=
    \sumk h^{T\Omega}_k\oft\,\mathcal{F}^{-1}\!\left\{W\A [\DDlnA\sij]
    \,a_{i|j}^k\right\}\!\oft
    \also
    \mathscr{A}_{j|i}\oft=
    \sumk h^{T\Omega}_k\oft\,\mathcal{F}^{-1}\!\left\{W\A [\DDlnA\sij]
    \,a_{j|i}^k\right\}\!\oft
    ,
  \end{align}
\end{subequations}
with which we are able to rewrite eqs~(\ref{blonki}) in the time
domain, via Plancherel's relation as applicable over the record
section of length~$T$. We have simultaneously extricated the tapers
from their seismograms, $\delta s_i^k\oft=\delta
[s_i^k\oft]^*=h^{T\Omega}_k\oft\,\delta s_i\oft$, and used the 
real-valuedness of the time-series to arrive at the recognizable
(see eq.~\ref{dchi})
form 
\begin{subequations}
  \begin{align}\label{eq:blonkt}
    \delta\chi\DD\PP=&
    \intot\left\{
    \sum_{i} \left[4\pi
      \sum_{j>i} \mathscr{S}_{i|j}\oft 
      \right]\delta s_i\oft
    +\sum_{j} \left[4\pi
      \sum_{i<j} \mathscr{S}_{j|i}\oft 
      \right]\delta s_j\oft 
    \right\}\dt
    ,
    \\
    \delta\chi\DD\A=&
    \intot\left\{
    \sum_{i} \left[4\pi
      \sum_{j>i} \mathscr{A}_{i|j}\oft 
      \right]\delta a_i\oft
    +\sum_{j} \left[4\pi
      \sum_{i<j} \mathscr{A}_{j|i}\oft 
      \right]\delta a_j\oft 
    \right\}\dt
    .
  \end{align}
\end{subequations}
The multitaper phase and amplitude adjoint sources are thus
\begin{align}\label{yolo}
  f_{\Phi,i}^{\dagger}(t)&=
  4\pi \sum_{j>i} \mathscr{S}_{i|j}\oftt
  \,\delta (\bx-\bx_i)
  \also
  f_{\Phi,j}^{\dagger}(t)=
  4\pi \sum_{i<j} \mathscr{S}_{j|i}\oftt 
  \,\delta (\bx-\bx_j)
  ,
  \\\label{polo}
  f_{A,i}^{\dagger}(t)&=
  4\pi \sum_{j>i} \mathscr{A}_{i|j}\oftt
  \,\delta (\bx-\bx_i)
  \also
  f_{A,j}^{\dagger}(t)=
  4\pi \sum_{i<j} \mathscr{A}_{j|i}\oftt
  \,\delta (\bx-\bx_j)
  .
\end{align}
It is gratifying to discover that eq.~(\ref{yolo}) is indeed the
multitaper generalization of eqs~(\ref{eq:adj_CC_DD}), which
themselves generalized eq.~(\ref{eq:adj_CC}). For its part,
eq.~(\ref{polo}), ultimately, is the double-difference multitaper
generalization of the amplitude-tomography adjoint source derived by,
e.g., \cite{Tromp+2005}, but which we have not given any further
consideration in the main text.

Finally, we return to the issue of symmetry in the differential
amplitude estimation, which can be restored by replacing all separate
summations of the type $\sum_i\sum_{j>i}$ and $\sum_j\sum_{i<j}$ that
appear in the expressions above by a common $\sum_{ij}$.

With these expressions we are finally able to define the multitaper
misfit sensitivity kernels along the same lines as eqs~(\ref{kp})
and~(\ref{kp2}), one for the phase and one for the amplitude
measurements, both in terms of the velocity perturbation --- we are
(yet) not concerned with variations in intrinsic attenuation
\cite[e.g.,][]{Tian+2009,Tian+2011,Zhu+2013}.

\subsection{Numerical experiment}

\begin{figure*}
  \rotatebox{-90}{\includegraphics[height=0.45\columnwidth]{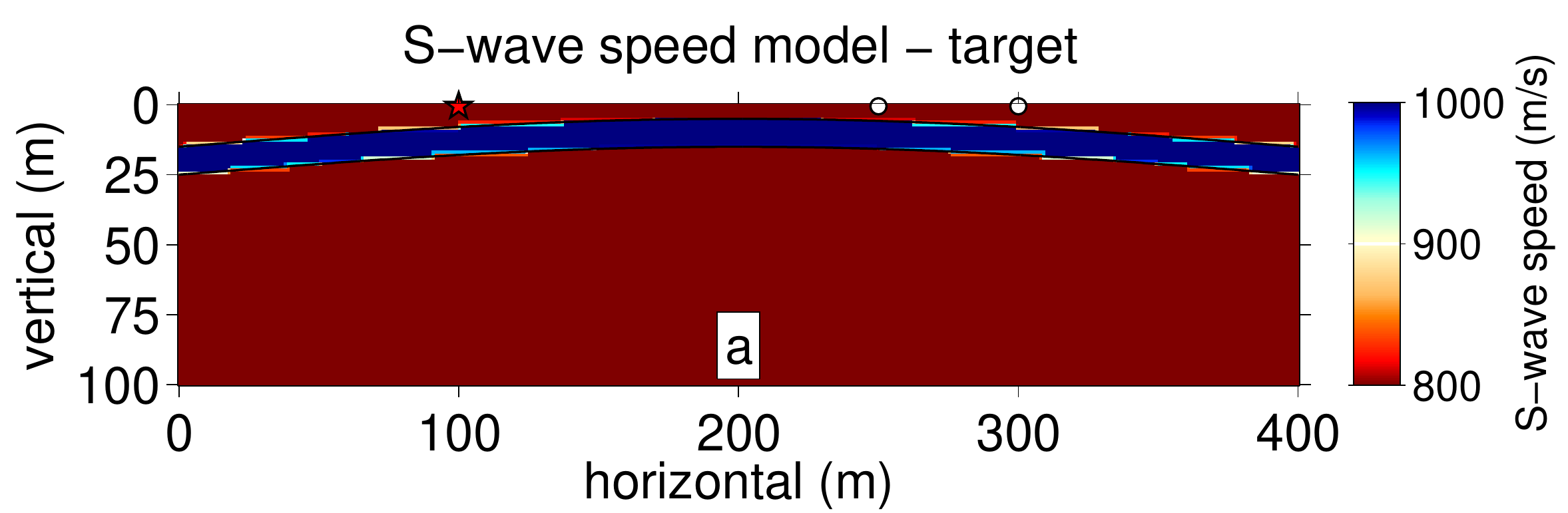}}\hspace{2em}
  \rotatebox{-90}{\includegraphics[height=0.45\columnwidth]{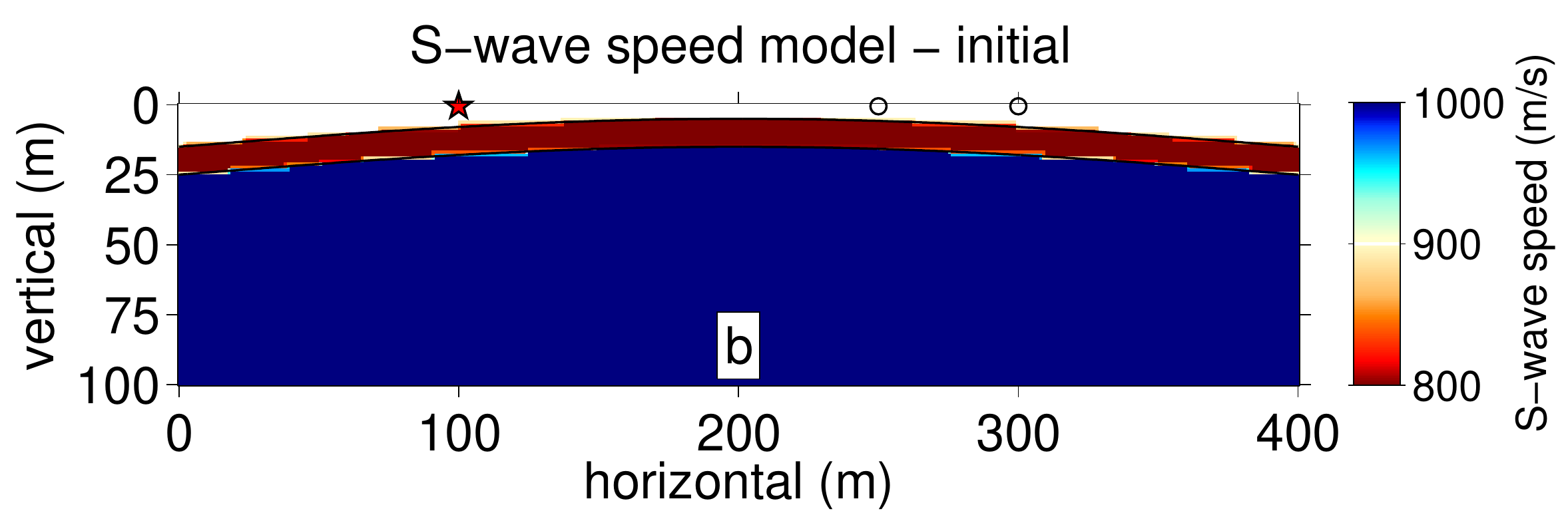}}
  \caption[Multitaper experiment: model configuration]{\label{fig:Sin_model}
    Experiment with multitaper phase and amplitude
    measurements. (a)~Target model. (b)~Initial model. The star
    is the source, the open circles depict two stations.}
\end{figure*}

Fig.~\ref{fig:Sin_model}a shows a simple two-dimensional vertical
\textit{S}-wave speed model with an anomalous layer that divides the
model into three segments: a top portion with an \textit{S}-wave speed
of 800~m/s, a middle with a speed of 1000~m/s, and a bottom with a
speed of 800~m/s. The initial model, shown in
Fig.~\ref{fig:Sin_model}b, possesses the same tripartite geometry but
the wavespeeds have different values: at the top 900~m/s, in the
middle 800~m/s, and at the bottom 1000~m/s.

Synthetics were computed used one point-source Ricker wavelet with
40~Hz dominant frequency located at 0.5~m depth, at a horizontal
distance of 100~m from the left edge of the model. Two receivers were
placed at the surface in the positions shown in
Fig.~\ref{fig:Sin_model}. 

\begin{figure*}
  {\includegraphics[width=0.48\textwidth]{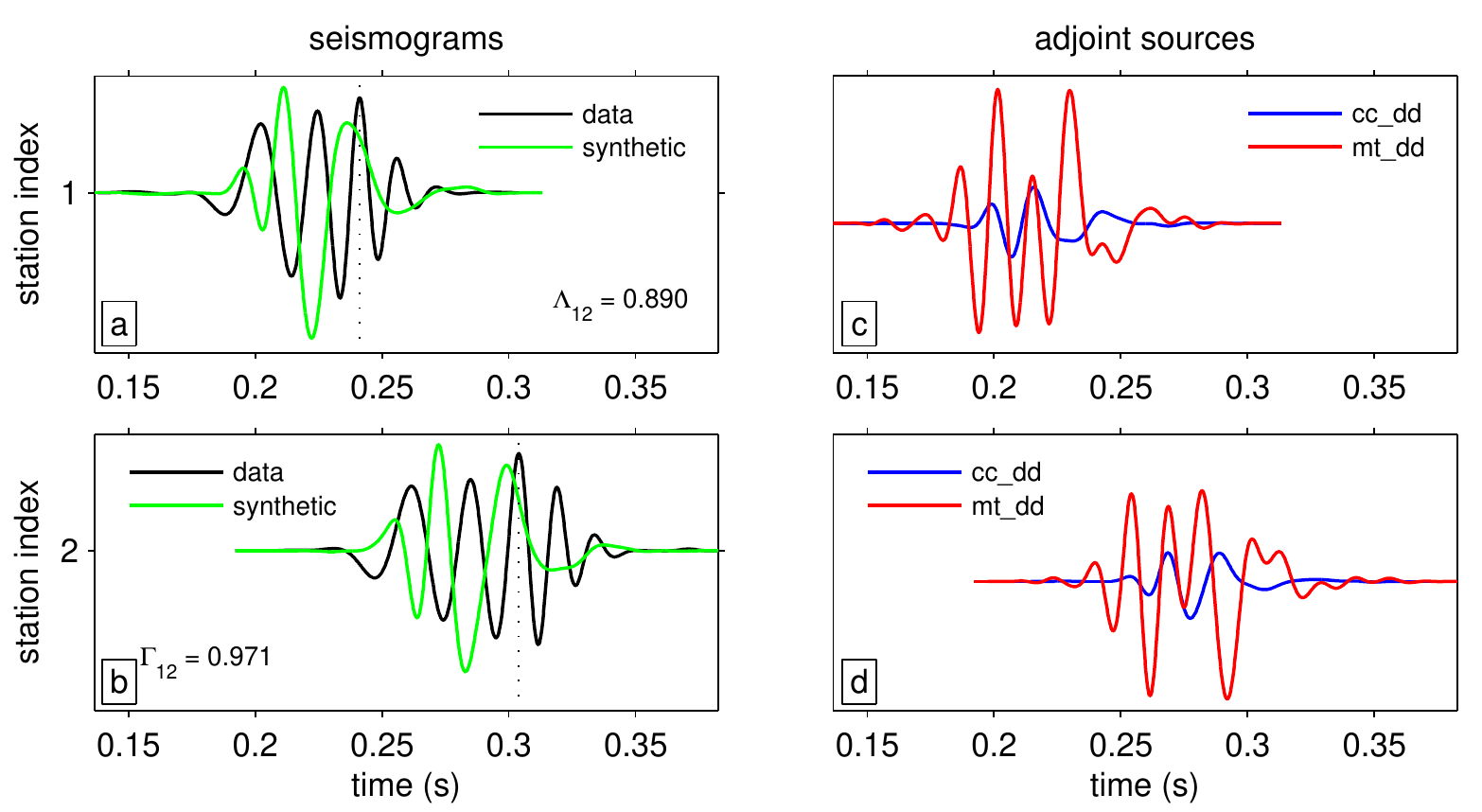}}\hspace{2em}
  {\includegraphics[width=0.48\textwidth]{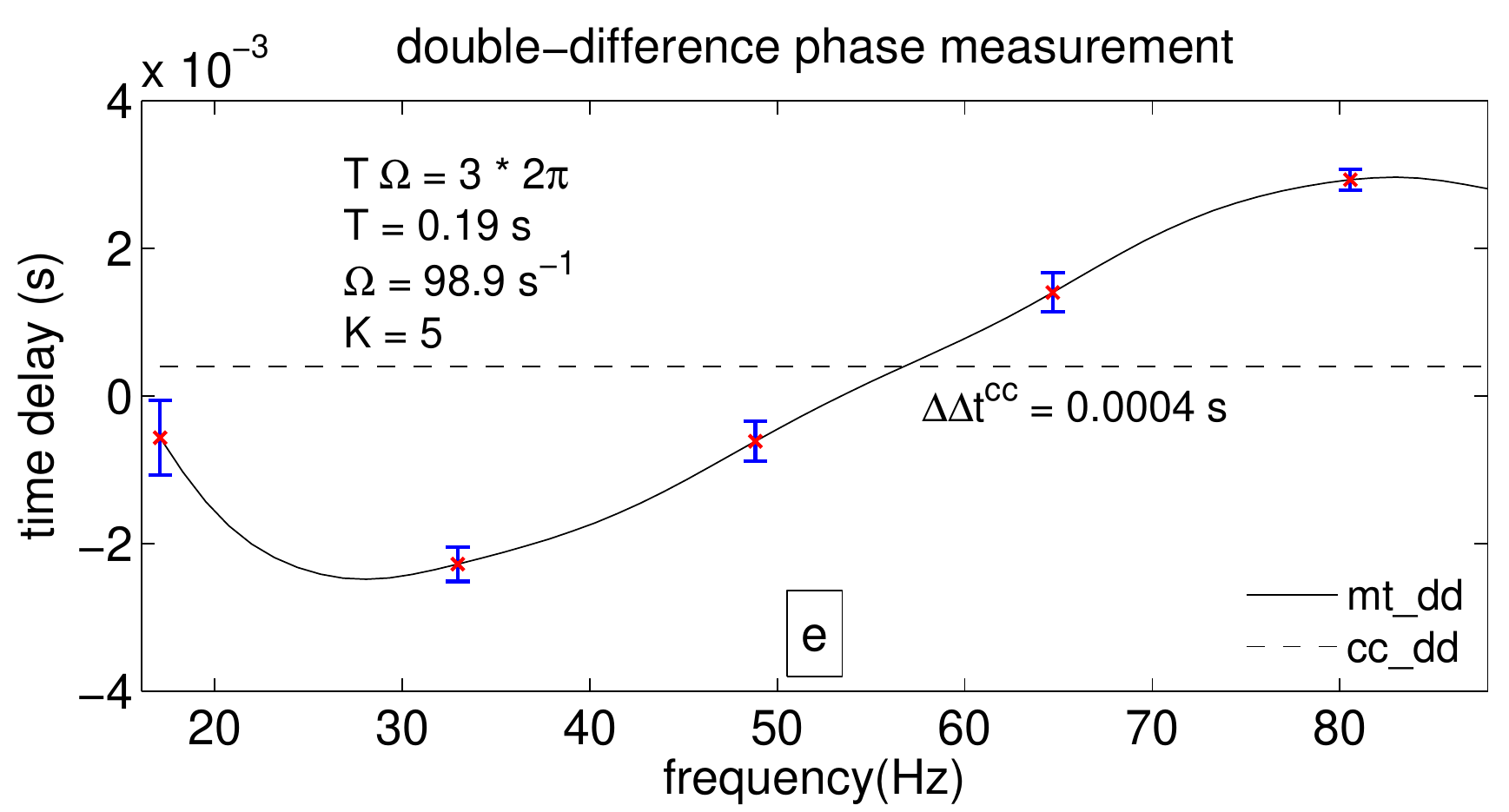}}
  \caption[Multitaper experiment: data and
    synthetics]{\label{fig:Sin_data} Experiment with multitaper phase
    and amplitude measurements. (a--b)~Data and
    synthetics. (c--d)~Adjoint sources. (e)~Double-difference phase
    measurements. The solid line are the multitaper estimates, with
    standard errors estimated using jackknifing. The dashed line is
    the cross-correlation value.}
\end{figure*}

\begin{figure*}
  \rotatebox{-90}{\includegraphics[height=0.45\columnwidth]{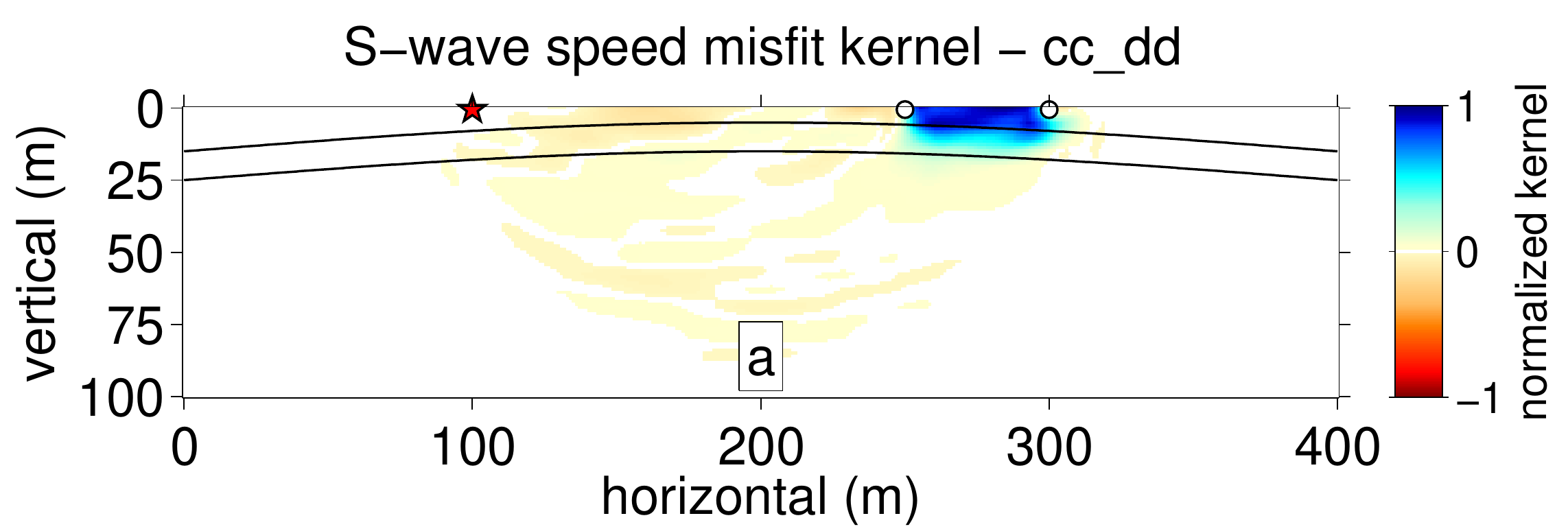}}\hspace{2em}
  \rotatebox{-90}{\includegraphics[height=0.45\columnwidth]{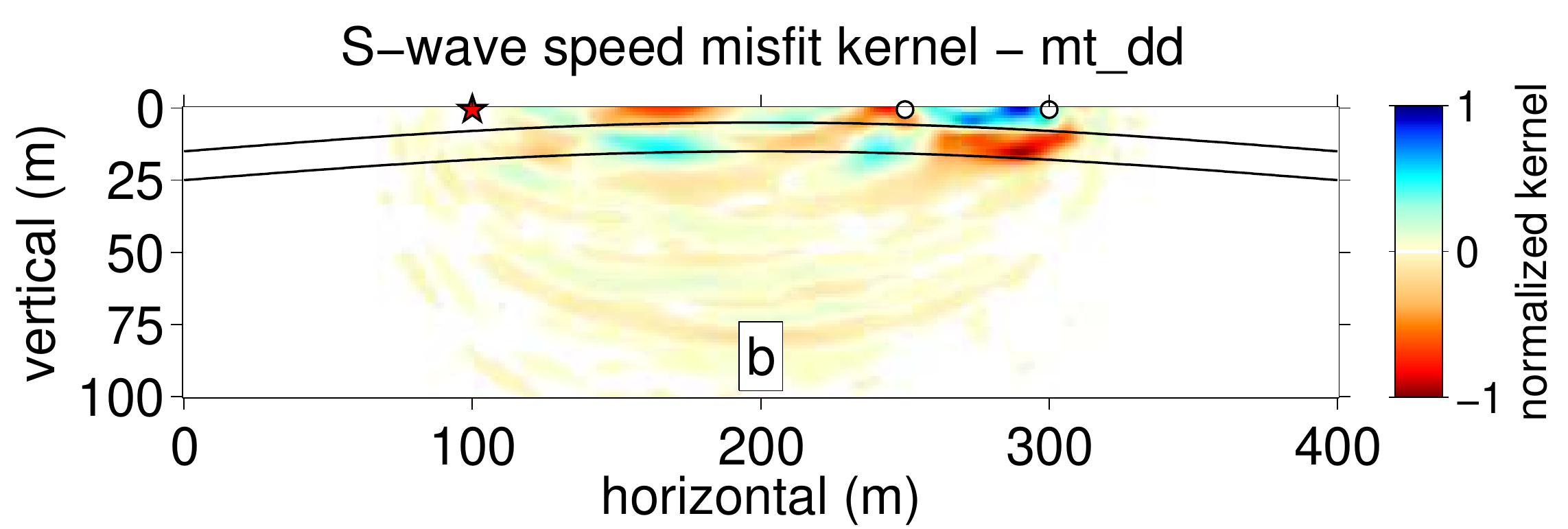}}
  \caption[Multitaper experiment: misfit sensitivity
    kernel]{\label{fig:Sin_kernel} Misfit sensitivity kernel for the
    (a)~double-difference of cross-correlation traveltimes and for the
    (b)~double-difference of multitaper phase measurements. The star
    is the source, the open circles two stations.}
\end{figure*}

Fig.~\ref{fig:Sin_data} shows the observed and synthetic records
(a--b), the adjoint sources (c--d), and the double-difference
frequency-dependent multitaper phase measurement
(e). Fig.~\ref{fig:Sin_kernel} compares the misfit sensitivity kernel
of the double-difference time-domain cross-correlation measurement to
the misfit sensitivity kernel of the double-difference multitaper
phase measurement, which shows some additional structure.

\end{document}